\documentclass[12pt,aps,prb,preprint,groupedaddress,showpacs]{revtex4}
\usepackage{graphicx,amssymb,bm,makeidx}
\usepackage{amsfonts}
\usepackage{amsmath,amssymb}

\newcommand{\I}{{\rm i}}

\begin{document}

\title{Magnetic fluctuations and self--energy effects in two--dimensional itinerant systems with van Hove singularity of electronic spectrum}
\author{P. A. Igoshev$^{a,b}$, V. Yu. Irkhin$^{a}$, and A. A. Katanin$^{a,b}$
}
\affiliation{$^{a}$Institute of Metal Physics, 620990, Kovalevskaya str. 18,
Ekaterinburg, Russia\\
$^{b}$Max-Planck Institute f\"ur Festk\"orperforschung, 70569, Stuttgart,
Germany}

\begin{abstract}
We investigate a competition of tendencies towards ferromagnetic and incommensurate order in two--dimensional fermionic systems within functional renormalization group technique using temperature as a scale parameter. We assume that the Fermi surface (FS) is substantially curved, lies in the vicinity of van Hove singularity points and perform an account of the self--energy corrections.  It is shown that the character of magnetic fluctuations is strongly asymmetric with respect to the Fermi level position relatively to van Hove singularity (VHS). For the Fermi level above VHS we find at low temperatures dominant incommensurate magnetic fluctuations, while below VHS level we find indications for ferromagnetic ground state. In agreement with the Mermin--Wagner theorem, at finite temperatures and in small magnetic fields we obtain rather small magnetization, which appears to be a power--law function of magnetic field. It is found that the FS curvature is slightly increased by correlation effects, and the renormalized bandwidth decreases at sufficiently low temperatures.
\end{abstract}

\maketitle

\section{Introduction}

During last two decades, the problem of magnetic fluctuations in itinerant--electron compounds attracts substantial interest in connection with physics of layered systems. In copper--oxide high--temperature superconductors the importance of magnetic fluctuations is seen from the proximity of antiferromagnetically ordered region to the region of superconducting state with pronounced incommensurate magnetic fluctuations\cite{La,Keimer}. Another impressive example of peculiar magnetic properties of low--dimensional systems is provided by the family of layered strontium ruthenates, which are pure metallic and incidentally have a very complex magnetic behavior.

Neutron scattering \cite{Sidis} of the one--layer paramagnetic metal Sr$_{2}$RuO$_{4}$ (which at low temperatures $T\lesssim 1.5$ K becomes an unconventional, most likely $p$--wave, superconductor\cite{Ru_review,Sc_review}) reveals an incommensurate character of magnetic fluctuation spectrum, with the largest contribution corresponding to the wave vector $\mathbf{Q}_{s}\sim (2/3)(\pi,\pi)$ and smaller one with the wave vector $\mathbf{Q}_{l}\sim (1/3)(\pi,\pi)$. It was argued\cite{Mazin} that incommensurate short--wave magnetic fluctuations characterized by the wavevector $\mathbf{Q}_{s}$ are expected to originate from $\alpha $ and $\beta $ bands (produced by $d_{xz}$ an $d_{yz}$ orbitals and generating quasi--one--dimensional sheets of FS), while the long--wave magnetic fluctuations with the wavevector $\mathbf{Q}_{l}$ originate most likely from the $\gamma$ band, produced by $d_{xy}$ orbitals and generating quasi--two--dimensional sheet of FS. Moreover, magnetic properties of this compound can be controlled by the substitution effect: being doped by lantanium, Sr$_{2-y}$La$_{y}$RuO$_{4}$ exhibits enhancement of magnetic susceptibility. This enhancement  is likely related to the position of the Fermi level of $\gamma$ band: The fit of the tight--binding spectrum to experimentally determined Fermi surfaces allows \textit{to expect} the onset of the ferromagnetic order at $y=0.27$, related to the elevation of the Fermi level of the $\gamma$ band towards a van Hove singularity (VHS). However, no ferromagnetic order was found for this doping (Ref. \onlinecite{Kikugawa}), despite the VHS at the Fermi level, revealed by ARPES \cite{TransverseVH}.

Bilayer ruthenate compound Sr$_{3}$Ru$_{2}$O$_{7}$ also provides a number of interesting physical properties, in particular incommensurate magnetic fluctuations \cite{Capogna} and metamagnetism \cite{Multi-Ru}; the latter was assumed to be related to the proximity of Fermi level to VHS\cite{Binz,Yuki}. The effect of substitution of Sr by Ca in (Sr$_{1-x}$Ca$_{x}$)$_{3}$Ru$_{2}$O$_{7}$ was studied in Ref. \onlinecite{2_Ru_doped_Ca}. Alhtough the Wilson ratio approaches $\sim 700$ for $x=0.2$, and the system becomes nearly ferromagnetic in the temperature interval ($3\lesssim T\lesssim 10$ K), further lowering of temperature does not result in the long--range ferromagnetic ordering, but instead  forces the system to freeze into a spin--glass state. Therefore, an important problem related to the magnetic properties of metallic two--dimensional compounds is revealing the conditions, for which the ferromagnetic order, which is observed to be suppressed experimentally, is stable.

From the theoretical point of view, magnetic properties of layered systems are closely related to their electronic structure. It is justified by ARPES experiments and first principle calculations of above--mentioned compounds that the electronic spectrum is quasi--two--dimensional and can be fit using nearest ($t$) and next--nearest neighbour ($t^{\prime }$) hoppings. To investigate an interplay of different magnetic states in itinerant layered systems it is convenient to consider the two--dimensional (2D) one--band Hubbard model. Despite the fact that this model is strongly simplified as compared to the situation in real compounds, it catches an important dependence of ground--state ordering on the form of the electronic spectrum, in particular on the $t^{\prime }/t$ ratio and band filling. On the other hand, the electronic spectrum containing nearest and next--nearest neighbor hoppings retains VHS which is often present in the density of states (DOS) of real compounds.

%From the point of view of the Stoner theory, which is, in essence, the mean-field approach to ferromagnetism in the Hubbard model \cite{Penn}, 
A large enough value of DOS at the Fermi level, occuring due to VHS, leads to ferromagnetic ground state according to the Stoner theory\cite{Penn}. This approach, however, neglects magnetic fluctuations, in particular, of incommensurate type. A generalization of the Stoner theory to treat the spin spiral instability and N\'{e}el antiferromagnetism reveals that in a major part of the phase diagram the spiral phase is even more preferable than the ferromagnetic one \cite{Schulz,OurIC}. In particular, the phases with the wave vectors $\mathbf{Q}=(Q,Q)$ and $\mathbf{Q}=(Q,\pi)$ were found to compete strongly with the ferromagnetic phase\cite{OurIC}. The critical value of the Hubbard interaction $U_{\mathrm{c}}$ for the ferromagnetic instability is smallest in the region with the Fermi level below VHS, where the competition of ferromagnetic and $(Q,\pi)$ phase is observed. On the other hand, in the vicinity and well above VHS the ferromagnetic phase is to a great extent suppressed by the incommensurate $(Q,Q)$ phase. 

To investigate the stability of ferromagnetic order with respect to long--wave magnetic fluctuations in itinerant 2D electronic systems at moderate Coulomb interaction, the \textit{spin-fermion} model was recently used \cite{QS}. The static character of magnetic fluctuations in 2D systems with magnetically ordered ground state allowed to determine the boundaries of ferromagnetic region and the most important magnetic instability, competing with ferromagnetism. It was argued that the region of stability of ferromagnetism shrinks substantially, especially at VH filling, due to competition with incommensurate magnetic fluctuations, in qualitative agreement with mean--field\cite{OurIC} and variational results \cite{Okabe}. 

In the limit of large Coulomb repulsion the stability of saturated ferromagnetic ground state with respect to other magnetic structures was investigated starting from 1960s. In particular, Nagaoka has found that saturated ferromagnetic state is unstable with respect to the spin--wave excitation with the wave vector $\mathbf{Q}=(\pi,\pi)$, which implies that it competes with commensurate N\'{e}el antiferromagnetic order \cite{Nagaoka}. This result was later supplemented with variational principle results indicating that at least in part of the phase diagram the wave vector of the corresponding magnetic instability is incommensurate\cite{Okabe}. 

Dynamical mean field theory (DMFT) provides a tool for unbiased account of the local electronic correlation effects and determination of the favorable type of ground--state magnetic order for arbitrary interaction strength. Within DMFT the magnetic phase diagram for the Hubbard model was investigated on a Bethe \cite{Pruschke1,Pruschke2} and square \cite{Kotliar} lattices with nearest and next--nearest neighbor hoppings. For small $t'$ away from half filling and sufficiently large Coulomb interaction ferromagnetic, paramagnetic and incommensurate magnetic (the region where commensurate solutions of the DMFT equations do not converge) phases  were found. At large enough $t'\lesssim t/2$ the ferromagnetism is found to be dominating in a broad region of parameters \cite{Pruschke2}. 

The non--local electronic correlations for saturated ferromagnetic state can be treat in particular in the Kanamori's T--matrix approximation (which is exact in the limit of small carrier concentration), applied to this problem in Ref. \onlinecite{Hlubina}. However, this approach does not account for contributions from the other channels of electron interaction, which are important at not too small concentrations. The possibility of incommensurate magnetic ordering is also missed by this approximation.

%It was argued in Refs. \cite{Roth,Edwards2,Igarashi,Irkhin} that the spin-polaron instability of saturated ferromagnetic state competes with the abovediscussed spin-wave instabilities well away of half-filling. This instability can be interpreted as a result of dramatic strong renormalization of the electronic spectrum, which is by itself an important aspect of the problem of the stability of ferromagnetic phase. However, in these papers the problem was investigated from the side at large Coulomb interaction, where the self-energy effects are predominantly a result of electron-magnon interaction. In the case of not-strong Coulomb interaction resulting self-energy momentum dependence is a result of delicate cooperation of different channels renormalization and its enough investigation is missed up to now. 

To study the interplay of magnetic and electronic properties and their effect on the possibility of ferromagnetic instability, we focus here on the functional remormalization group (fRG) technique, which is a powerful tool for treatment of correlation effects (see Ref. \onlinecite{fRG} for the introduction in the application to the Hubbard model). In the pioneer study of Ref. \onlinecite{Zanchi}, the use of the Polchinski's form of fRG technique allowed to construct the phase diagram of the nearest--neighbor hopping Hubbard model ($\mu-T$ plane). The momentum cut--off scheme of the Wick--ordered version of fRG equations was employed to obtain the ground--state instability type at small $t^{\prime }/t$ in Ref. \onlinecite{Halboth}. Later the temperature \cite{Salmhofer,Kampf,Katanin} and Hubbard interaction \cite{Honerkamp} were used as flow parameters in one--particle irreducible (1PI) investigations of the $t-t'$ Hubbard model. Main result of these studies is that the small nearest--neighbor--hopping $(t^{\prime }/t\leq0.2)$ favors the antiferromagnetic instability at VH filling, moderate one $(0.2\leq t^{\prime }/t\leq 0.35)$, the $d$--wave supeconducting instability, rather large $t^{\prime }/t>0.35$ corresponds to ferromagnetic instability. Away from van Hove band filling the competition of antiferromagnetic and superconducting instabilities is obtained at small $t^{\prime }/t$, while at larger values $t^{\prime }/t\gtrsim 0.35$ the competition of ferromagnetic instability and $p$--wave superconductivity is observed \cite{Salmhofer,Kampf,Honerkamp}.

The essential shortcoming of the above reviewed approaches \cite{Salmhofer,Honerkamp,Halboth,Kampf} is that these do not consider the self--energy corrections to the electronic Green function. Hence, important non--trivial effects of renormalization of the electronic spectrum in the vicinity of magnetic phase transition (or in the regime with strong magnetic fluctuations) can be missed, provided that the Fermi level is near VHS. On the other hand, spin--dependent self--energy corrections provide the mechanism of the response to magnetic field, which is crucial in the context of invesigation of ferromagnetic instability. Note that only antiferromagnetic and superconducting phases were considered previously within the combination of fRG and mean--field approach (Ref. \onlinecite{Rohe}) and within the fRG approach in the symmetry broken phase \cite{Lauscher}. Considering ferromagnetic instability poses a problem of scale--dependent FS, which was theoretically elaborated only in the self--adjusting Polchinski and Wick--ordered schemes, proposed in Ref. \onlinecite{Salmhofer1}.

In this paper we present a study of the evolution of magnetic and electronic properties with decreasing temperature within 1PI fRG in zero and small finite magnetic field. We treat accurately the FS problem, including movement of projecting points and momentum dependence of the self--energy. The plan of the paper is the following. In Sec. \ref{model_diagram} we introduce the model and review earlier approaches to ferromagnetic instability at weak and intermediate coupling. In Sec. \ref{SE_fRG} we introduce the details of our novel fRG approach to take into account the fluctuation effects, retaining the electronic self--energy. In Sec. \ref{Results} we present and discuss the numerical results. In Sec. \ref{Conclusions} we present the conclusions. 

\section{The model and earlier approaches}

%\section{Model}
\label{model_diagram} %\subsection{Model}
We consider the Hubbard model with the action 
\begin{equation}
\mathcal{S}=\beta \sum_{k\sigma }(-\mathrm{i}\nu_{n}+\epsilon_{\mathbf{k}%
}-\mu -h\sigma )c_{k\sigma }^{+}c_{k\sigma }+\frac{\beta
U}{4N}\sum_{k_{1}k_{2}k_{3}k_{4}\sigma \sigma ^{\prime }}c_{k_{1}\sigma
}^{+}c_{k_{2}\sigma ^{\prime }}^{+}c_{k_{3}\sigma ^{\prime }}c_{k_{4}\sigma
}\delta_{k_{1}+k_{2},k_{3}+k_{4}},  
\label{Action}
\end{equation}%
where $\beta=1/T$ is inverse temperature, $N$ is the number of lattice sites, $h$ is a magnetic field directed along $z$ axis, measured in units of Bohr magneton, $\epsilon_{\mathbf{k}}$ is an electronic spectrum, $\mu$ is the chemical potential, $U $ is the Hubbard on--site interaction, $\delta$ is the Kronecker $\delta$--symbol. The sums in Eq. (\ref{Action}) are taken over 4--vectors $k=(\mathrm{i}\nu_{n},\mathbf{k}),$ where $\nu_{n}=\pi(2n+1)T$ are the fermionic Matsubara frequencies ($n\in \mathbb{Z}$). We consider electronic dispersion on the square lattice 
\begin{equation}
\epsilon_{\mathbf{k}}=-2t(\cos k_{x}+\cos k_{x})+4t^{\prime }(\cos
k_{x}\cos k_{y}+1),  
\label{bare_ek}
\end{equation}%
where $t,t^{\prime}>0$ are hopping parameters. Such a form of the spectrum corresponds to VHS at zero energy. We fix the ratio $t^{\prime}/t=0.45$.

\begin{figure}
\includegraphics{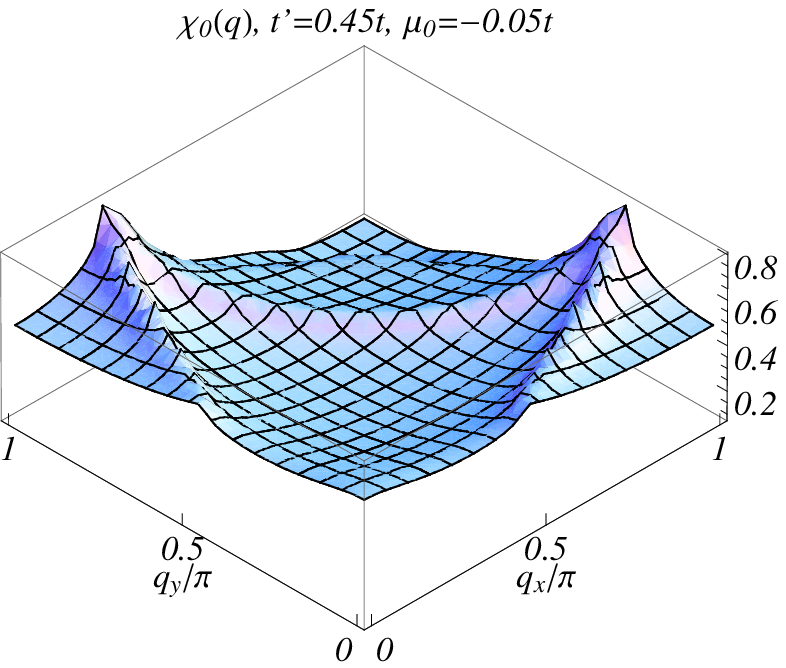}\includegraphics{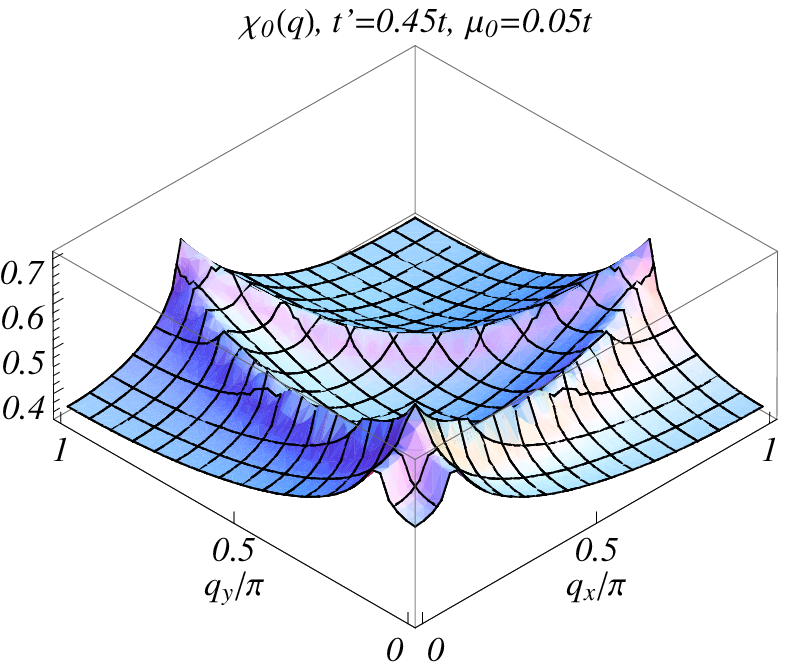}
\caption{Momentum dependence of the static magnetic susceptibility of free electrons ${\chi^0_{\mathbf{q}}(\omega=0)}$, ${t^{\prime}=0.45t}$: a) Fermi level below VH filling (${\mu=-0.05t}$); b) Fermi level above VH filling (${\mu=0.05t}$)}
\label{Sus}
\end{figure}

%As it was discussed in the Introduction, besides conventional ferro- and antiferromagnetic orders incommensurate magnetic structures are good candidates to order of the ground state in the case of sufficienly large $t'$ (ratio $t'/t\lesssim1/2$). 
To gain insight into possible types of magnetic order, we consider the magnetic susceptibility $\chi_{\mathbf{q}}(\omega)$  in paramagnetic state \cite{Moriya},
\begin{equation}
\chi_{\mathbf{q}}^{-1}(\omega)=\phi_{\mathbf{q}}^{-1}(\omega)-U,
\label{RPA}
\end{equation}
where $\phi_{\mathbf{q}}(\omega )$ is the susceptibility, irreducible in particle--hole channel. The paramagnetic state is unstable with respect to the formation of incommensurate state with the wave vector $\mathbf{Q}$, provided that the maximum of irreducible susceptibility $\phi_{\mathbf{q}}(0)$ is at $\mathbf{q=Q}$ and $\chi_{\mathbf{Q}}(0)$ diverges. A critical interaction for stability of such an incommensurate magnetic state $U_{\rm c}=\phi_{\mathbf{Q}}^{-1}(0)$. For the sake of simplicity let us consider the random phase approximation (RPA) where $\phi_{\mathbf{q}}(\omega)$ coincides with the non--interacting spin susceptibility 
\begin{equation}
\chi^0_{\mathbf{q}}(\I\omega_n)=-\frac{T}{N}\sum_{\mathbf{k}\nu
_{n}}G_{\mathbf{k}}^{0}(\mathrm{i}\nu_{n})G_{\mathbf{k}+\mathbf{q}}^{0}(%
\mathrm{i}\nu_{n}+\mathrm{i}\omega_{n}),  
\label{Sus0}
\end{equation}
where $G_{\mathbf{k}}^{0}(\mathrm{i}\nu_{n})=(\mathrm{i}\nu_{n}-\epsilon_{\mathbf{k}}+\mu)^{-1}$ is the non--interacting electronic Green function. In Fig. \ref{Sus} we present zero--temperature momentum profile of static non--interacting magnetic susceptibility $\chi^0_{\mathbf{q}}(\omega=0)$. Below VH filling the competition between incommensurate magnetic structure with large wave vector and a variety of incommensurate instabilities with small magnetic wave vectors is observed (see Fig. \ref{Sus}a). On the other hand, above VH filling, the ferromagnetic instability competes with the long--wave incommensurate magnetic instability \cite{QS}(see Fig. \ref{Sus}b). This consideration shows that for large enough $t'$ the tendency to incommensurate magnetic ordering originates from the geometry of the FS. 

The competition of ferromagnetism and incommensurate magnetic ordering beyond mean--field approximation was discussed within the quasistatic approach (QSA) in Ref. \onlinecite{QS}. Under the assumption of the ground state to be ferromagnetically ordered, it was found that magnetic fluctuations can reconstruct the momentum dependence of the susceptibility only at finite electronic interaction, stabilizing the ferromagnetic ground state with respect to incommensurate magnetic fluctuations. Above the critical value of the interaction, the maximum of $\phi_{\mathbf{q}}(0)$ is reached at $\mathbf{q=0}$. The corresponding result for the boundary separating the ferromagnetic and $(Q,Q)$ incommensurate instabilities is presented below in Fig. \ref{Summary}. 

The effort of direct account of the electron correlation effects was performed in temperature--flow fRG study\cite{Salmhofer,Kampf} ($\Sigma=0$--fRG). While the problem of ferromagnetism formation was investigated in zero magnetic field, the self--energy renormalization was neglected, which makes the FS scale independent and the chemical potential not renormalized. Formal consequences of this assumption are discussed in Sec. \ref{Relation_prev}.

\section{Functional renormalization group with self--energy corrections}

\label{SE_fRG}Present study is an extension of the study of Ref. \onlinecite{Kampf}. Contrary to previous approaches, we (i) do not neglect the self--energy corrections $\Sigma$ to the electronic Green's function $G$ and account for their momentum dependence, which results in the renormalization of spectrum parameters $t$ and $t^{\prime }$ and moving FS, (ii) partially take into account momentum dependence of the vertex inside the patches, which allows to account for the scale dependence of the Fermi surfaces $\mathcal{F}_{\sigma }(s)$, corresponding to different spin projections $\sigma=\pm1$.

\subsection{fRG equations}

\label{fRG_equations} We use the 1PI fRG equations \cite{fRG} in the truncation of Ref. \onlinecite{Ward} 
\begin{equation}
\dot{\Gamma}=\left. \Gamma \ast \frac{d}{ds}(G_sG_s)\ast \Gamma \right\vert_{\mathrm{pp}}+\left. \Gamma \ast \frac{d}{ds}(G_sG_s)\ast \Gamma \right\vert_{\mathrm{ph}}+\left. \Gamma \ast \frac{d}{ds}(G_sG_s)\ast \Gamma \right\vert_{\mathrm{ph1}},  \label{fRG_symbol1}
\end{equation}
\begin{equation}
\dot{\Sigma}=\Gamma \ast S,  
\label{fRG_symbol2}
\end{equation}
where $\Gamma $ is the 1PI 4--vertex function (we refer to it as a vertex below), pp, ph, and ph1 denote particle--particle and two independent particle--hole channels, $\Sigma $ is the self--energy, $S=-G_s^2dG^{-1}_{0,s}/ds$ is the single--scale propagator, $G_{0,s}$ and $G_s$ are appropriately rescaled non--interacting and interacting Green's functions, momentum arguments and spin indices are omitted for brevity. The asteriscs denote the summation over internal 4--momentum and spin indices corresponding to specified channel, dots denote the derivatives, taken with respect to the scale parameter $s$.

If FM order parameter or magnetic field are directed along $z$ axis, the SU(2) symmetry is broken. The remaining axial symmetry results in the self--energy which is diagonal in spin indices $\sigma,\sigma^{\prime}$ 
\begin{equation}
\Sigma_{\sigma \sigma ^{\prime }}(k)=\Sigma_{k\sigma}\delta_{\sigma\sigma ^{\prime}},
\end{equation}%
and the vertex function components are nonzero provided that $\sigma_1+\sigma_2=\sigma_3+\sigma_4$. We consider the vertex component $\Gamma_{\sigma_{1}\sigma_{2};\sigma _{1}\sigma _{2}}$ (for brevity we denote it as $\Gamma _{\sigma _{1}\sigma _{2}}$), the others can be obtained using exact relation 
\begin{equation}
\Gamma_{\sigma_{1}\sigma_{2};\sigma_{3}\sigma_{4}}(k_{1},k_{2};k_{3},k_{4})=-\Gamma_{\sigma _{2}\sigma _{1};\sigma
_{3}\sigma_{4}}(k_{2},k_{1};k_{3},k_{4})=\Gamma_{\sigma_{2}\sigma_{1};\sigma_{4}\sigma_{3}}(k_{2},k_{1};k_{4},k_{3}).
\end{equation}
%Another useful property which is a consequence of the symmetry given by the time-reversal, combined with $\sigma\rightarrow-\sigma$ transformation, is
Another useful property, which is a consequence of the symmetry with respect to the combination of time--reversal transformation and $\sigma\rightarrow-\sigma$ transformation, is  
\begin{equation}
\Gamma_{\sigma _{1}\sigma _{2};\sigma _{3}\sigma_{4}}(k_{1},k_{2};k_{3},k_{4})=\bar{\Gamma}_{\sigma_{3}\sigma_{4};\sigma
_{1}\sigma _{2}}(k_{3},k_{4};k_{1},k_{2}).
\end{equation}

Neglecting the frequency dependence of the vertex and self--energy the fRG equations can be written in the following explicit form:
\begin{multline}
\label{DGamma}
\dot{\Gamma}_{\sigma _{1}\sigma_{2}}(\mathbf{k}_{1},\mathbf{k}_{2};\mathbf{k%
}_{3},\mathbf{k}_{4})=\\
=(1-\delta_{\sigma_{1}\sigma _{2}}/2)\frac1{N}\sum_{\mathbf{p}}\Gamma _{\sigma
_{1}\sigma _{2}}(\mathbf{k}_{1},\mathbf{k}_{2};\mathbf{p},\mathbf{k}_{1}+%
\mathbf{k}_{2}-\mathbf{p})\mathcal{L}_{\sigma _{1}\sigma _{2}}^{\mathrm{pp}}(%
\mathbf{p},\mathbf{k}_{1}+\mathbf{k}_{2}-\mathbf{p})\Gamma _{\sigma
_{1}\sigma _{2}}(\mathbf{p},\mathbf{k}_{1}+\mathbf{k}_{2}-\mathbf{p};\mathbf{%
k}_{3},\mathbf{k}_{4}) \\
-\frac1{N}\sum_{\mathbf{p}\sigma }\Gamma _{\sigma _{1}\sigma }(\mathbf{k}_{1},\mathbf{%
p};\mathbf{k}_{3},\mathbf{k}_{1}-\mathbf{k}_{3}+\mathbf{p})\mathcal{L}%
_{\sigma \sigma }^{\mathrm{ph}}(\mathbf{p},\mathbf{k}_{1}-\mathbf{k}_{3}+%
\mathbf{p})\Gamma _{\sigma \sigma _{2}}(\mathbf{k}_{1}-\mathbf{k}_{3}+%
\mathbf{p},\mathbf{k}_{2};\mathbf{p},\mathbf{k}_{4}) \\
+\delta_{\sigma _{1}\sigma _{2}}\frac1{N}\sum_{\mathbf{p}\sigma }\Gamma _{\sigma
_{1}\sigma }(\mathbf{k}_{1},\mathbf{k}_{2}-\mathbf{k}_{3}+\mathbf{p};\mathbf{%
k}_{4},\mathbf{p})\mathcal{L}_{\sigma \sigma }^{\mathrm{ph}}(\mathbf{p},%
\mathbf{k}_{2}-\mathbf{k}_{3}+\mathbf{p})\Gamma _{\sigma \sigma _{2}}(%
\mathbf{p},\mathbf{k}_{2};\mathbf{k}_{2}-\mathbf{k}_{3}+\mathbf{p},\mathbf{k}%
_{3})] \\
+(1-\delta _{\sigma _{1}\sigma _{2}})\frac1{N}\sum_{\mathbf{p}}\Gamma _{\sigma
_{1}\sigma _{2}}(\mathbf{k}_{1},\mathbf{k}_{2}-\mathbf{k}_{3}+\mathbf{p};%
\mathbf{p},\mathbf{k}_{4})\mathcal{L}_{\sigma _{1}\sigma _{2}}^{\mathrm{ph}}(%
\mathbf{p},\mathbf{k}_{2}-\mathbf{k}_{3}+\mathbf{p})\Gamma _{\sigma
_{1}\sigma _{2}}(\mathbf{p},\mathbf{k}_{2};\mathbf{k}_{3},\mathbf{k}_{2}-%
\mathbf{k}_{3}+\mathbf{p})
\end{multline}
\begin{equation}
\dot{\Sigma}_{\mathbf{k}\sigma }=\frac{1}{2N}\sum_{\mathbf{p}\sigma ^{\prime
}}\Gamma _{\sigma \sigma ^{\prime }}(\mathbf{k},\mathbf{p};\mathbf{k},%
\mathbf{p})\left[ f_{\mathbf{p}\sigma ^{\prime }}+(2\varepsilon _{\mathbf{p}%
\sigma ^{\prime }}-\Sigma _{\mathbf{p}\sigma ^{\prime }})f_{\mathbf{p}\sigma
^{\prime }}^{\prime }\right] -\dot{\mu}\sum_{\mathbf{p}\sigma ^{\prime
}}\Gamma _{\sigma \sigma ^{\prime }}(\mathbf{k},\mathbf{p};\mathbf{k},%
\mathbf{p})f_{\mathbf{p}\sigma ^{\prime }}^{\prime }-\Sigma _{\mathbf{k}%
\sigma }/2,  \label{DSigma}
\end{equation}%
where 
\begin{multline}
\mathcal{L}_{\sigma \sigma ^{\prime }}^{\mathrm{pp}}(\mathbf{k},\mathbf{k}%
^{\prime })=\left[ \frac{\varepsilon _{\mathbf{k}\sigma }f_{\mathbf{k}\sigma
}^{\prime }+\varepsilon _{\mathbf{k}^{\prime }\sigma ^{\prime }}f_{\mathbf{k}%
^{\prime }\sigma ^{\prime }}^{\prime }}{\varepsilon _{\mathbf{k}\sigma
}+\varepsilon _{\mathbf{k}^{\prime }\sigma ^{\prime }}}+\frac{(\dot{\Sigma}_{%
\mathbf{k}\sigma }-\dot{\mu})f_{\mathbf{k}\sigma }^{\prime }+(\dot{\Sigma}_{%
\mathbf{k}^{\prime }\sigma ^{\prime }}-\dot{\mu})f_{\mathbf{k}^{\prime
}\sigma ^{\prime }}^{\prime }}{\varepsilon _{\mathbf{k}\sigma }+\varepsilon
_{\mathbf{k}^{\prime }\sigma ^{\prime }}}\right.  \label{Lpp} \\
-\left. \frac{(f_{\mathbf{k}\sigma }+f_{\mathbf{k}^{\prime }\sigma ^{\prime
}}-1)(\dot{\Sigma}_{\mathbf{k}\sigma }+\dot{\Sigma}_{\mathbf{k}^{\prime
}\sigma ^{\prime }}-2\dot{\mu})}{(\varepsilon _{\mathbf{k}\sigma
}+\varepsilon _{\mathbf{k}^{\prime }\sigma ^{\prime }})^{2}}\right],
\end{multline}%
\begin{equation}
\label{Lph}
\mathcal{L}_{\sigma \sigma ^{\prime }}^{\mathrm{ph}}(\mathbf{k},\mathbf{k}%
^{\prime })=-\left[ \frac{\varepsilon _{\mathbf{k}\sigma }f_{\mathbf{k}%
\sigma }^{\prime }-\varepsilon _{\mathbf{k}^{\prime }\sigma ^{\prime }}f_{%
\mathbf{k}^{\prime }\sigma ^{\prime }}^{\prime }}{\varepsilon _{\mathbf{k}%
\sigma }-\varepsilon _{\mathbf{k}^{\prime }\sigma ^{\prime }}}+\frac{(\dot{%
\Sigma}_{\mathbf{k}\sigma }-\dot{\mu})f_{\mathbf{k}\sigma }^{\prime }-(\dot{%
\Sigma}_{\mathbf{k}^{\prime }\sigma ^{\prime }}-\dot{\mu})f_{\mathbf{k}%
^{\prime }\sigma ^{\prime }}^{\prime }}{\varepsilon _{\mathbf{k}\sigma
}-\varepsilon _{\mathbf{k}^{\prime }\sigma ^{\prime }}}-\frac{(f_{\mathbf{k}%
\sigma }-f_{\mathbf{k}^{\prime }\sigma ^{\prime }})(\dot{\Sigma}_{\mathbf{k}%
\sigma }-\dot{\Sigma}_{\mathbf{k}^{\prime }\sigma ^{\prime }})}{(\varepsilon
_{\mathbf{k}\sigma }-\varepsilon _{\mathbf{k}^{\prime }\sigma ^{\prime
}})^{2}}\right].  
\end{equation}%
We have introduced the renormalized electronic spectrum 
\begin{equation}
\varepsilon _{\mathbf{k}\sigma }=\epsilon_{\mathbf{k}}-\mu-h\sigma+\Sigma_{\mathbf{k}\sigma},  
\label{renorm_e}
\end{equation}%
and $f_{\mathbf{p}\sigma }\equiv f(\varepsilon _{\mathbf{p}\sigma })=\frac{1%
}{2}(1-\tanh (\beta \varepsilon _{\mathbf{p}\sigma }/2)).$ 

Below we consider temperature cutoff \cite{Salmhofer} with $s=\log(t/T)$.  The single--scale propagator has a form 
\begin{equation}
S_{k\sigma }^{T}=-G_{k\sigma }^{2}\left( \frac{\mathrm{i}\nu _{n}-\epsilon_{\mathbf{k}}-h\sigma+\mu}{2T^{1/2}}+T^{1/2}\frac{d\mu }{dT}\right)
\label{S_T}
\end{equation}
where $G_{k\sigma}=(\mathrm{i}\nu_{n}-\varepsilon _{\mathbf{k}\sigma })^{-1}$. In the present study we choose temperature independent bare chemical potential ($d\mu/dT=0$), although the temperature dependence of $\mu$ can be adjusted, e.~g., to keep the number of particles fixed\cite{Comment}. The equations (\ref{DGamma}) and (\ref{DSigma}) are basic for 1PI fRG approach since they can, in principle, determine the renormalization of the electronic spectrum and interaction at any temperature. This system of equations should be supplemented by initial conditions at $s=-\infty(T=+\infty )$. In the infinite--temperature limit correlation effects are absent, and one obtains for the self--energy $\Sigma_{\mathbf{k}\sigma }=U/2$ and interaction $\Gamma_{\sigma_{1}\sigma_{2}}(k_{1},k_{2};k_{3},k_{4})=U(1-\delta_{\sigma_{1}\sigma_{2}}).$

The fRG equations (\ref{DGamma}),(\ref{DSigma}) form the system of ordinary \textit{functional} differential equations. To solve them numerically, one has to introduce some approximation procedure to reduce the considered system to a finite system of ordinary equations. Below we present the procedure of numerical solution in details. 

\subsection{The self--energy ansatz and Fermi surface}

The first step is the use of common patching scheme \cite{Salmhofer} to avoid dealing with system of functional equations. However, we do not neglect the momentum dependence of the self--energy (and vertices) inside the patches, in particular Eq. (\ref{DSigma}) is fully employed. 
\label{self-energy_ansatz} 
We assume that the self--energy has the form 
\begin{equation}
\label{se_ansatz}
\Sigma_{\mathbf{k}\sigma }=-2\delta t_{\sigma }(\cos k_{x}+\cos
k_{y})+4\delta t_{\sigma }^{\prime }\cos k_{x}\cos k_{y}+\Sigma_{\sigma}.
\end{equation}%
where the parameters $\delta t_{\sigma },\delta t_{\sigma }^{\prime}$ and $\Sigma_{\sigma}$ are determined as follows. While solving numerically the system of fRG equations, at each step we calculate the values of $\dot{\Sigma}_{\mathbf{k}\sigma }$ on two sets of ($\sigma$--dependent) projecting points (PPs, see for details Appendix). For considered $\mathbf{k}\in$ PPs the linear regression procedure is used,
\begin{equation}
\dot{\Sigma}_{\mathbf{k}\sigma }\rightarrow -2\delta \dot{t}_{\sigma }(\cos
k_{x}+\cos k_{y})+4\delta \dot{t}_{\sigma }^{\prime }\cos k_{x}\cos k_{y}+%
\dot{\Sigma}_{\sigma }.
\end{equation}%
In this way we determine the flow of unknown quantities $\delta t_{\sigma},\delta t_{\sigma }^{\prime }$ and $\Sigma_{\sigma}$ in Eqs. (\ref{DGamma}),(\ref{DSigma}).

Such a choice efficiently reduces the number of variables and retains VHS points of the renormalized spectrum. Note that $\Sigma_{\sigma}$ does not depend on $\mathbf{k}$ and renormalizes the chemical potential $\mu ;\delta t_{\sigma }$ and $\delta t_{\sigma}^{\prime}$ contribute to the change of momentum dependence of the spectrum ($\delta t_{\sigma }$ corresponds to the bandwidth renormalization).

Therefore, it is convenient to represent the renormalized spectrum (\ref{renorm_e}) in the form 
\begin{equation}
\varepsilon_{\mathbf{k}\sigma }=-2t_{\mathrm{eff,\sigma }}(\cos k_{x}+\cos
k_{x})+4t_{\mathrm{eff,\sigma }}^{\prime }(\cos k_{x}\cos k_{y}+1)-\mu _{%
\mathrm{eff,\sigma }},
\end{equation}
where
\begin{equation}
\mu_{\mathrm{eff,}\sigma }=\mu-\Sigma_{\sigma }+4\delta t_{\sigma
}^{\prime }+h\sigma ,\;t_{\mathrm{eff,\sigma }}=t+\delta t_{\sigma }\;t_{\mathrm{eff,\sigma }}^{\prime }=t^{\prime }+\delta t_{\sigma }^{\prime}.
\label{mu_eff}
\end{equation}%
The scale--dependent effective chemical potential $\mu _{\mathrm{eff,}\sigma}$ results in Fermi surface, calculated with the renormalized spectrum parameters $t_{\mathrm{eff,\sigma }},t_{\mathrm{eff,\sigma }}^{\prime }$. Applying present method one should be careful in determining the geometry of the FS. If $t_{\mathrm{eff,\sigma }}^{\prime }/t_{\mathrm{eff,\sigma }}<1/2,$ the bottom of the band is at the energy $w_{\sigma }=-4t_{\mathrm{eff,\sigma }}+8t_{\mathrm{eff,\sigma }}^{\prime }$ and the FS is singly connected; however, if $t_{\mathrm{eff,\sigma }}^{\prime }/t_{\mathrm{eff,\sigma }}\geq 1/2$ the bottom of the band is zero. In this case if $\mu_{\mathrm{eff,\sigma }}\geq w_{\sigma },$ the FS is singly connected and the patching scheme of Ref. \onlinecite{Zanchi} can be used, while for $\mu _{\mathrm{eff,\sigma}}<w_{\sigma}$, the Fermi surface consists of two disconnected parts and the patching scheme should be chosen differently.

Despite that the self--energy renormalization are included in the present study, we neglect incoherent contributions to the Green functions. Due to this renormalization, the actual Fermi level $\mu_{\mathrm{eff}}$ is determined by the combination of the bare spectrum and the self--energy parameters (\ref{mu_eff}) at the end of the flow.

\subsection{The vertex ansatz}

\label{vertex_ansatz} 
The vertex function $\Gamma $ is represented by its values at the current FS. Since we always trace the renormalization of $\Sigma _{\mathbf{k}\sigma }$ we have to take into account moving of the Fermi surface during fRG flow. 
PPs of the current FS are changing during the flow and the discrete (projected) vertex function derivative acquires an additional contribution corresponding to this movement, 
\begin{equation}
\frac{d\Gamma}{ds}=\frac{\partial\Gamma}{\partial s}+\frac{\partial \Gamma}{\partial k_{\mathrm{PP}}}\frac{dk_{\mathrm{PP}}}{ds}.
\end{equation}
We denote symbolically the derivatives with respect to PPs ($k_{PP}$) as ${\partial }/{\partial k_{\mathrm{PP}}}.$

To take into account this momentum dependence of $\Gamma$ we assume that, apart from the position of external legs in certain patches, the vertex function depends on momenta $\mathbf{k}_{i}$ through the renormalized energies $\varepsilon _{\mathbf{k}_{i}\sigma }$ and linearize the latter dependence. Let the momenta of external legs $\mathbf{k}_{1}^{\mathrm{c}},\mathbf{k}_{2}^{\mathrm{c}},\mathbf{k}_{3}^{\mathrm{c}}$ be on the current FS and consider vertex with all momenta $\mathbf{k}_{i}$ belonging to the main set of PPs and three vertices with two momenta $\mathbf{k}_{i}$ belonging to the main set of PPs (see Appendix), and one belonging to the auxiliary set. Therefore we have 4 possibilities for the choice of $\mathbf{k}_{1},\mathbf{k}_{2},\mathbf{k}_{3}$, and obtain a system of four linear equations (we use $\varepsilon _{\mathbf{k}_{i}^{\mathrm{c}}\sigma_{i}}=0$) 
\begin{equation}
\label{v_ansatz}
\Gamma_{\sigma \sigma ^{\prime }}(\mathbf{k}_{1}^{\mathrm{c}},\mathbf{k}_{2}^{\mathrm{c}};\mathbf{k}_{3}^{\mathrm{c}}))+\sum_{i}\partial _{i}\Gamma_{\sigma\sigma^{\prime }}(\mathbf{k}_{1}^{\mathrm{c}},\mathbf{k}_{2}^{%
\mathrm{c}};\mathbf{k}_{3}^{\mathrm{c}}))\varepsilon _{\mathbf{k}_{i}\sigma_{i}}=\Gamma _{\sigma \sigma ^{\prime }}(\mathbf{k}_{1},\mathbf{k}_{2};\mathbf{k}_{3}),  
\end{equation}
with unknown quantities $\Gamma _{\sigma \sigma ^{\prime }}(\mathbf{k}_{1}^{\mathrm{c}},\mathbf{k}_{2}^{\mathrm{c}};\mathbf{k}_{3}^{\mathrm{c}}))$ and $\partial _{i}\Gamma _{\sigma \sigma ^{\prime }}(\mathbf{k}_{1}^{\mathrm{c}},\mathbf{k}_{2}^{\mathrm{c}};\mathbf{k}_{3}^{\mathrm{c}})$. Solving the system (\ref{v_ansatz}) one determines the vertex on the current FS.

\subsection{Relation to previous approaches}
\label{Relation_prev}
In zero magnetic field, neglecting the momentum dependence of the self--energy (i. e. with $\delta t=\delta t^{\prime}=0$), one can demand the chemical potential to absorb all the corrections to the self--energy in Eq. (\ref{DGamma}), which is possible since the electronic spectrum enters the Eq. (\ref{DGamma}) through $\varepsilon_{\mathbf{k}\sigma}$ only. On the other hand, Eq. (\ref{DSigma}), being reformulated with $\mu ^{\prime}=\mu-\Sigma$, reduces to $\dot{\mu}^{\prime }=0$. Hence, in this case, $\mu $ is replaced by a constant $\mu ^{\prime}$. The important consequence of this approximation is that FS is fixed within this ansatz, since it is determined solely by the scale--independent parameter $\mu ^{\prime}$.

If the field is non--zero, the self--energy corrections cannot be absorbed into the chemical potential even when the renormalization of the hopping parameters is neglected. This, in turn, results in moving of spin FSs during the flow, so that projected vertex acquires a contribution from this moving. As described in detail in Sect. \ref{vertex_ansatz}, we account for this contribution in the self--consistent numerical scheme of treatment of Eqs. (\ref{DGamma}), (\ref{DSigma}), and (\ref{v_ansatz}).

\section{Temperature dependence of the renormalized parameters}

\label{Results}

In this Section we present and discuss numerical results of the present fRG approach accounting for the self--energy corrections in zero (Sec. \ref{h=0}) and finite magnetic field (Sec. \ref{h<>0}). Afterwards, in Sec. \ref{Phase_diagram} we present our results for the phase diagram. 

\subsection{Results in zero magnetic field}

\label{h=0} In this subsection we consider the results of fRG calculations in the spin symmetric phase for $U/t=3$ and 4. We choose the bare chemical potential $\mu$ in such a way that at the end of the flow the renormalized position of the Fermi level $\mu_{\mathrm{eff}}$, determined by Eq. (\ref{mu_eff}), lies in the vicinity of VHS. It is clear physically and verified numerically in our investigation that the magnetic response is suppressed for the Fermi level well separated from VHS. 

\begin{figure}
\includegraphics[width=.5\textwidth]{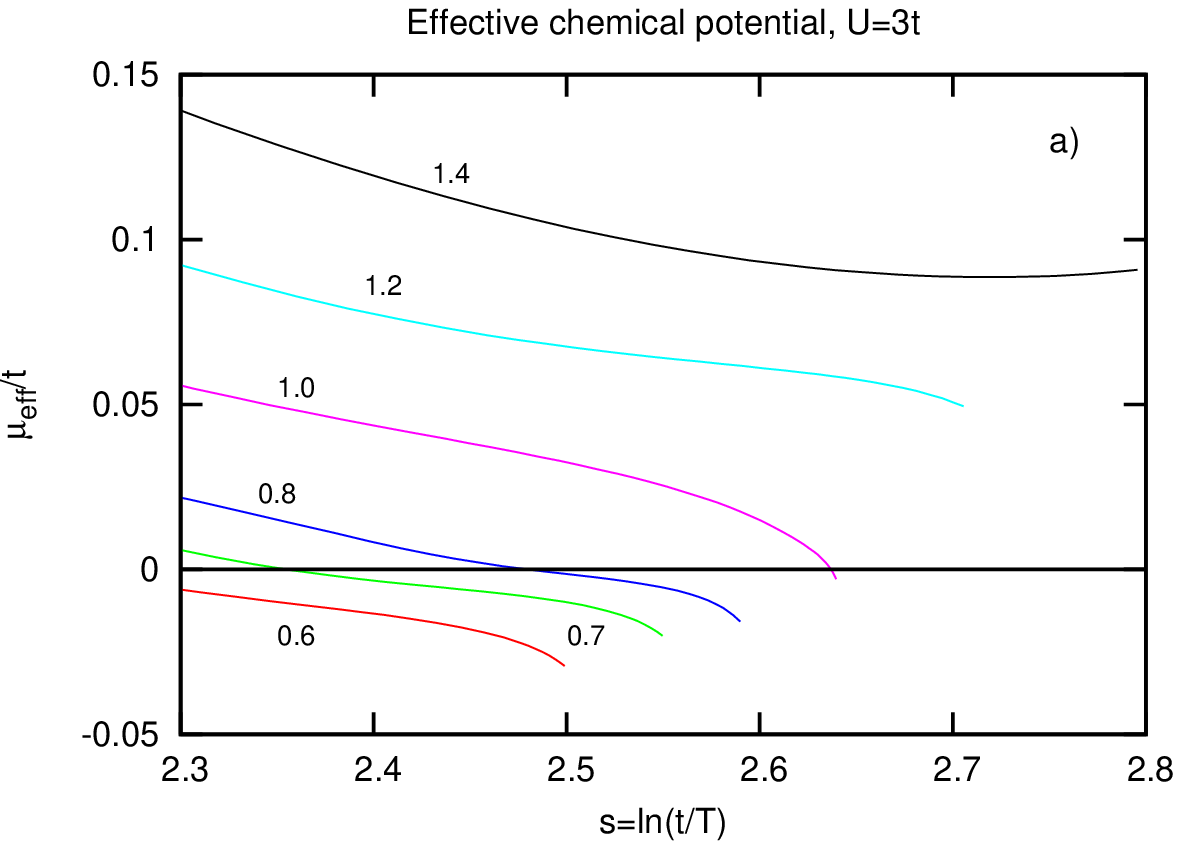}%
\includegraphics[width=.5\textwidth]{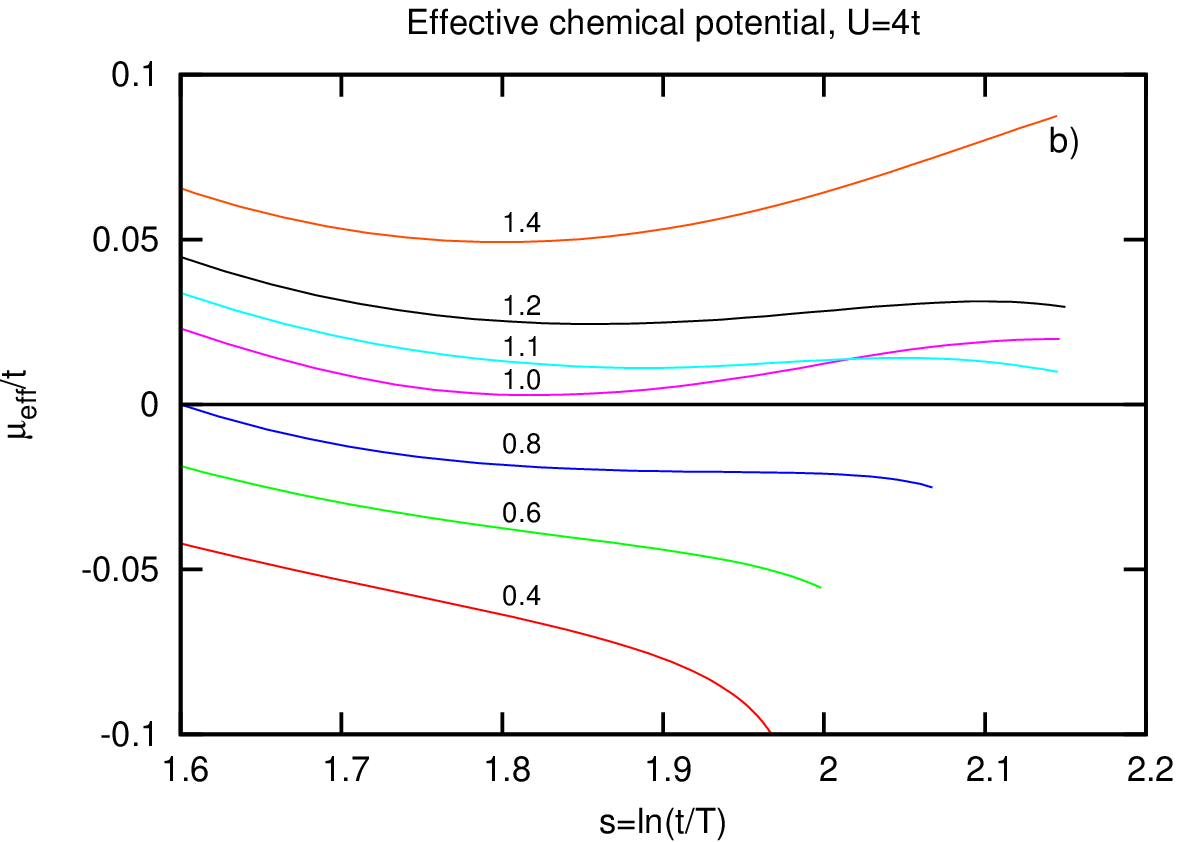}
\caption{Temperature dependence of the renormalized Fermi level $\mu_{\mathrm{eff}}$ at $h=0$: a) $U=3t$, b) $U=4t$. Chemical potential values $\bar\mu$ (see text) are shown by numbers near the plots.}
\label{fermi_fig}
\end{figure}

Starting from an infinite--temperature limit the position of the Fermi level first tends to increase from its high--temperature Hartree value $\mu-U/2$ to almost low--temperature Hartree value $\bar\mu=\mu-U\int_{-4t+8t'}^{\bar\mu}\rho(\varepsilon) d\varepsilon$, $\rho$ being the bare DOS. At even lower temperatures $\mu$ decreases due to correlation effects. The obtained low--temperature scale dependence of $\mu_{\mathrm{eff}}$ for different bare chemical potentials $\mu$ is shown in Fig. \ref{fermi_fig}a ($U/t=3$) and Fig. \ref{fermi_fig}b ($U/t=4$). We stop the flow, when the effective interaction becomes too large ($\gtrsim2.5W$, where $W=8t$ is bare bandwidth). The corresponding scale $s_{\rm min}$ yields the minimal temperature which is available within the flow, $T_{\rm min}=t\exp(-s_{\rm min})$. In the following we parametrize the initial chemical potential of flow by the Hartree Fermi level $\bar\mu$. 
%The common feature of the cases $U=3t$ and $U=4t$ is that in the case of Fermi level above VHS ($\mu_{\rm eff}>0$) $T_{\rm min}$  are substantially lower as compared to the case $\mu_{\rm eff}<0$. (This is also seen from the results for the maximal vertex, see Figs. \ref{vertex_3},\ref{vertex_4}). Hovewer, 

The cases $U=3t$ and $U=4t$ are somewhat different due to the absence of the region of saturation of $\mu_{\rm eff}(s)$ dependence for $U=3t$, although the dependence $\mu_{\rm eff}(s)$ becomes weak at the end of the flow. Below we consider the renormalized Fermi level $\mu_{\rm eff}$ in the saturation region (if it exists) as a representative parameter which characterizes the flow, since the renormalization of other parameters of the electronic spectrum is not too strong. The charge response is slightly suppressed in the vicinity of van Hove filling, for the case $\mu_{\rm eff}<0$ more than for $\mu_{\rm eff}>0$. Non--monotonous behavior of $\mu_{\mathrm{eff}}(s)$ slightly above VHS ($\bar\mu=1.0t$) for $U=4t$, which consequences are discussed below, is worthy of notice. 

%In the case $\mu_{\mathrm{eff}}<0$ (Fermi level is below van Hove singularities), the charge response is slightly suppressed (relatively to the case $\mu_{\rm eff}>0$) for $U=3t$ and the dependence $\mu_{\rm eff}(s)$ does not come up to the region of saturation (see Fig. \ref{fermi_fig}a); for $U=4t $ this relative suppression is absent (charge response is uniform both above and below VHS), and the scale dependence of the Fermi level acquires a flat featureless region while $\mu_{\rm eff}(s)$ approaches the VHS at the end of the flow ($\mu=0.8t$). We consider the case $\mu_{\rm eff}>0$. In the case $U=3t$ we find that temperatures of entering critical region are substantially lower with respect to the case $\mu_{\rm eff}<0$.  For $U=4t$ (see Fig. \ref{fermi_fig}b) $\mu_{\mathrm{eff}}(s)$ has a non-monotonic behavior with respect to $s$ (temperature) in the vicinity of VHS with possible merging of $\mu_{\mathrm{eff}}(s)$ plots for different chemical potentials $\mu$ ($\mu=1.0t,1.1t$). These pecularities of the scale dependence of the chemical potentials are discussed below. 
\begin{figure}
\includegraphics[width=.5\textwidth]{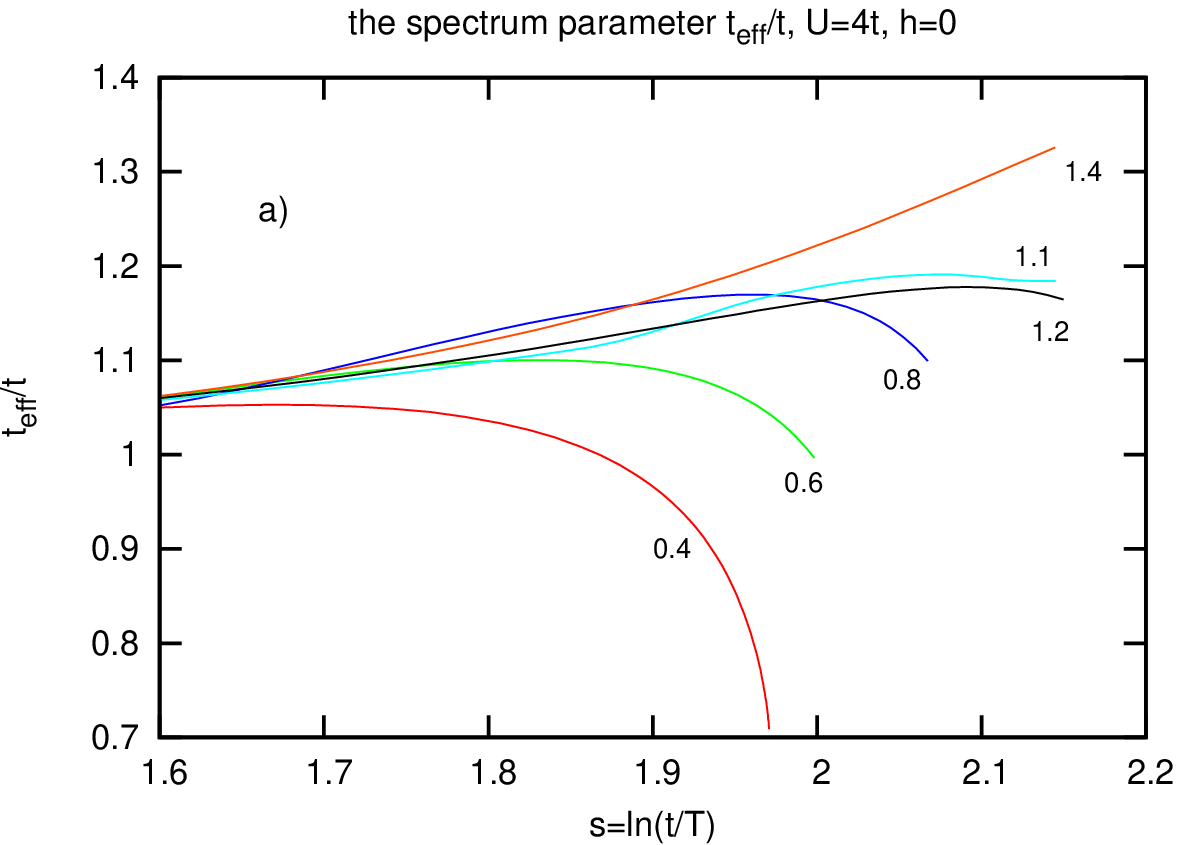}%
\includegraphics[width=.5\textwidth]{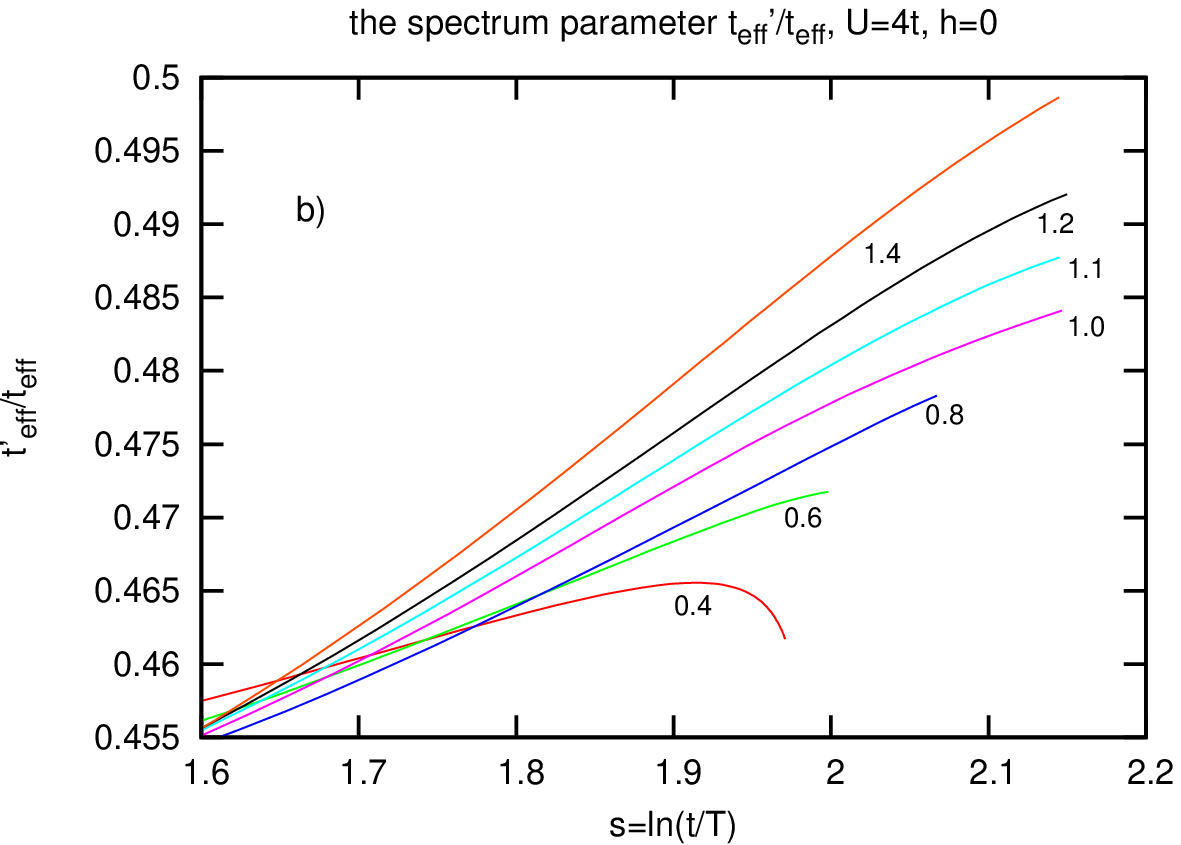}
\caption{Effective (renormalized) hopping parameters: a) $t_{\mathrm{eff}}/t$, b) $t_{\mathrm{eff}}^{\prime }/t_{\mathrm{eff}}$ for $U=4t$. Numbers correspond to the value of chemical potential $\bar\mu$}
\label{spectrum_fig}
\end{figure}

In our scheme the effective chemical potential renormalization has substantial contribution from $t$ and $t^{\prime }$ renormalizations (see Eq. (\ref{mu_eff})). In Fig. \ref{spectrum_fig} we present for instance $t_{\mathrm{eff}}/t$ and $t_{\mathrm{eff}}^{\prime }/t_{\mathrm{eff}}$ plots for $U=4t$. While the bandwidth ($t_{\mathrm{eff}}/t$) is somewhat reduced at the end of the flow well below VHS, indicating prominent correlation effects; for $\mu_{\rm eff}$ above VHS $t_{\rm eff}/t$ first increases with decreasing temperature (which implies that correlation effects in this regime are not substantial) and then decreases at lower temperatures, which is related to enhancement of the vertex in the vicinity of magnetic instability. On the other hand, the ratio $t^{\prime }/t$ is not strongly renormalized (this means that the self--energy effects do not change substantially the curvative of the Fermi surface) and monotonously increases towards the value 1/2 as $\bar\mu$ increases, which corresponds to flattening of the electronic dispersion. The complete treatment of the case $\bar\mu \geq 1.4t$, where $t_{\mathrm{eff}}^{\prime}/t_{\mathrm{eff}}$ exceeds $1/2$ during the flow, corresponding to a change of FS geometry (see for details Sec. \ref{self-energy_ansatz}), requires special consideration and is beyond the scope of the present study. 

The type of leading magnetic instability can be inferred from the behavior of vertices. Let us consider the scale profiles of maximal vertex. We consider maximal total vertex $\Gamma_{\uparrow \downarrow }^{\mathrm{max}}$ (the maximum is taken over all possible combinations of momenta) and maximal exchange vertex $\Gamma_{\uparrow\downarrow}^{\mathrm{max,E}}$ (the maximum is taken over all combinations of momenta with $\mathbf{k}_{2}=\mathbf{k}_{3}$). The ferromagnetic instability is accompanied by coincidence of $\Gamma_{\uparrow\downarrow}^{\mathrm{max}}$ and $\Gamma_{\uparrow \downarrow }^{\mathrm{max,E}}$ in the vicinity of transition point where both the values diverge, reflecting an instability of the paramagnetic state with respect to zero--momentum collective spin excitations. %However, since we know only partial momentum dependence of the vertex within the used patching procedure (incoming and outgoing momenta lie at the current Fermi level and have finite number of values), the application of this criterion to the specification of the type of leading instability is not fully exact. For instance, the ferromagnetic instability is undistinguishible from long-wave incommensurate magnetic instability, which corresponds to small wave vector, within this criterion: both these instabilities results in coincideness of $\Gamma_{\uparrow\downarrow}^{\mathrm{max}}$ and $\Gamma_{\uparrow\downarrow }^{\mathrm{max,E}}$: the difference $\Delta\Gamma_{\rm max}\equiv\Gamma_{\uparrow \downarrow }^{\mathrm{max}}-\Gamma_{\uparrow\downarrow }^{\mathrm{max,E}}$ vanishes, since maximums are taken over momentum in the projecting points only. 
Therefore, the criterion for commensurate magnetic fluctuations has the form 
\begin{equation}
\label{commensurability}
\Delta\Gamma_{\rm max}\equiv\Gamma_{\uparrow \downarrow }^{\mathrm{max}}-\Gamma_{\uparrow\downarrow}^{\mathrm{max,E}}=0
\end{equation}
and yields an information about the type of leading instability.

\begin{figure}
\includegraphics[width=0.5\textwidth]{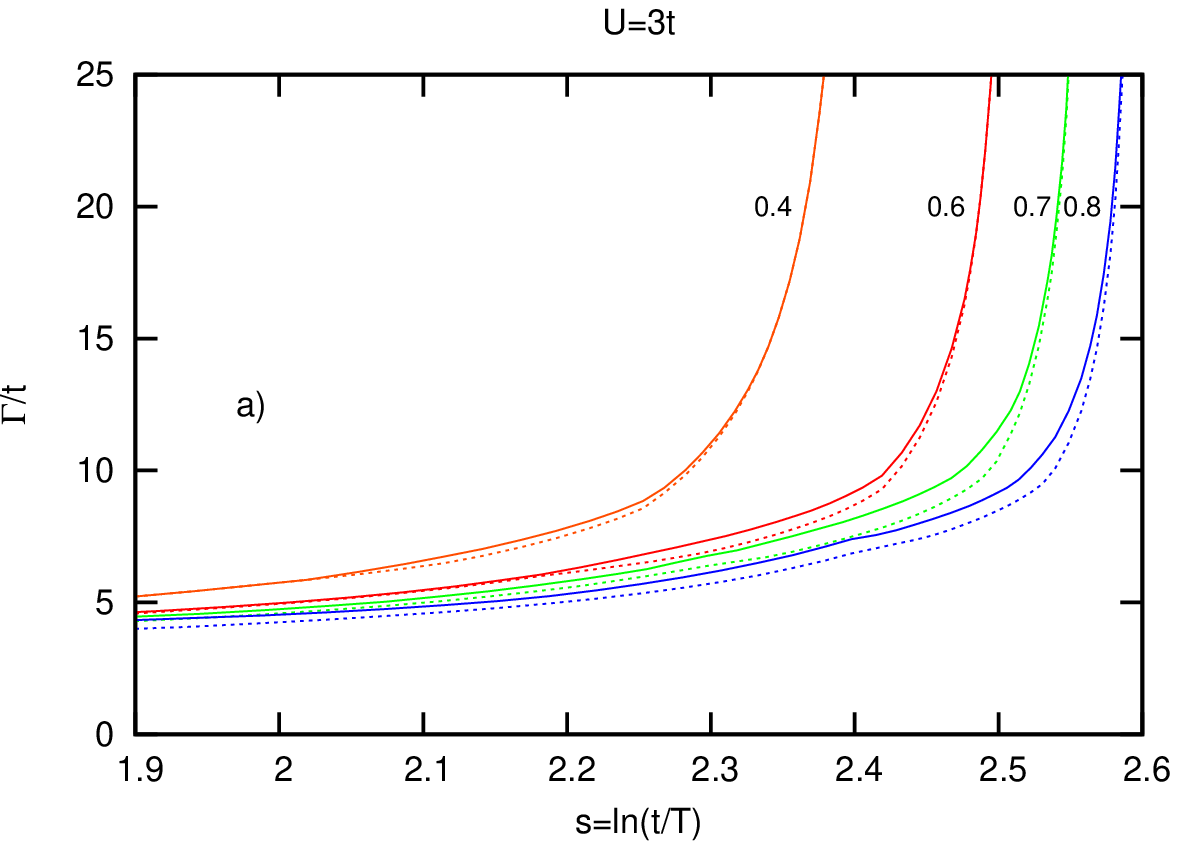}%
\includegraphics[width=0.5\textwidth]{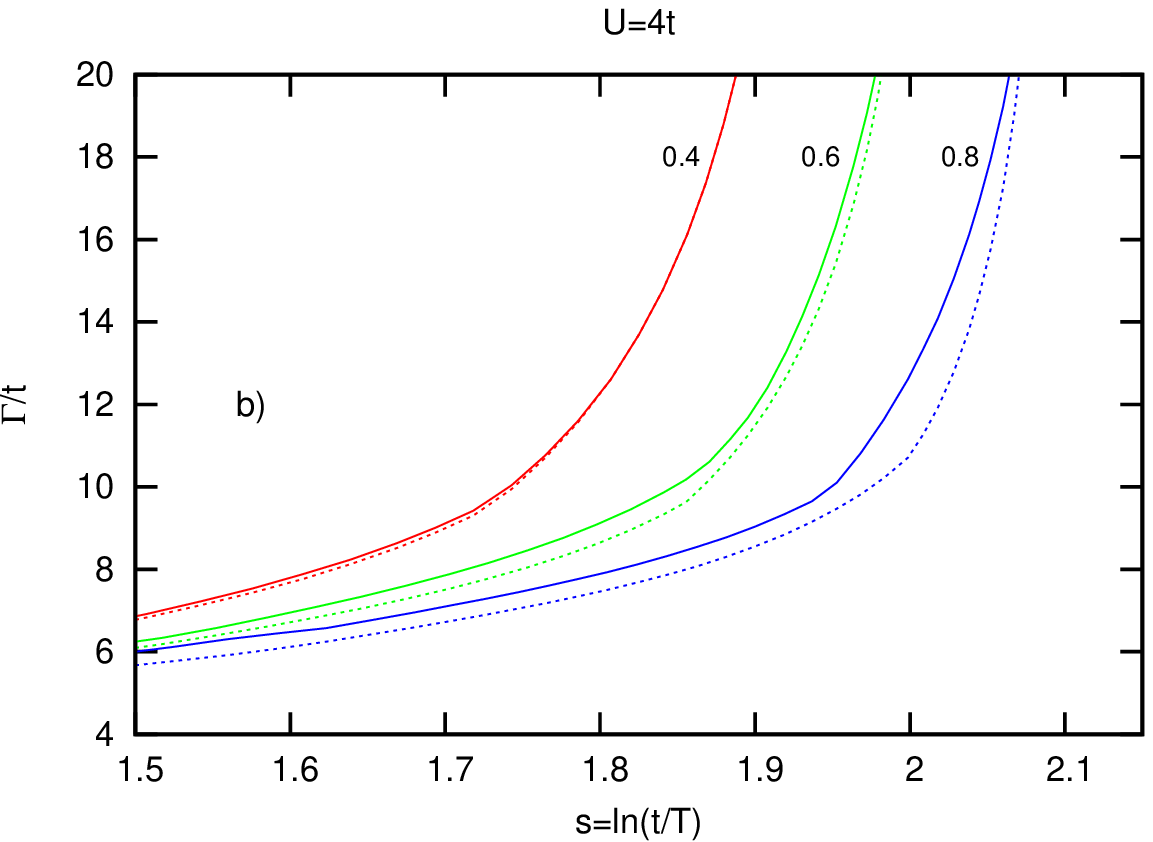}
\caption{The scale profiles of $\Gamma^{\mathrm{max}}_{\uparrow\downarrow}$ (solid lines) and $\Gamma^{\mathrm{max, E}}_{\uparrow\downarrow}$ (dashed lines), see text, in the case $h=0$ a) $U=3t$, b) $U=4t$ for different choices of the chemical potential $\bar\mu$ (shown by numbers) yielding Fermi level below VHS ($\mu_{\mathrm{eff}}<0$)}
\label{vertex_3}
\end{figure}
\begin{figure}
\includegraphics[width=0.5\textwidth]{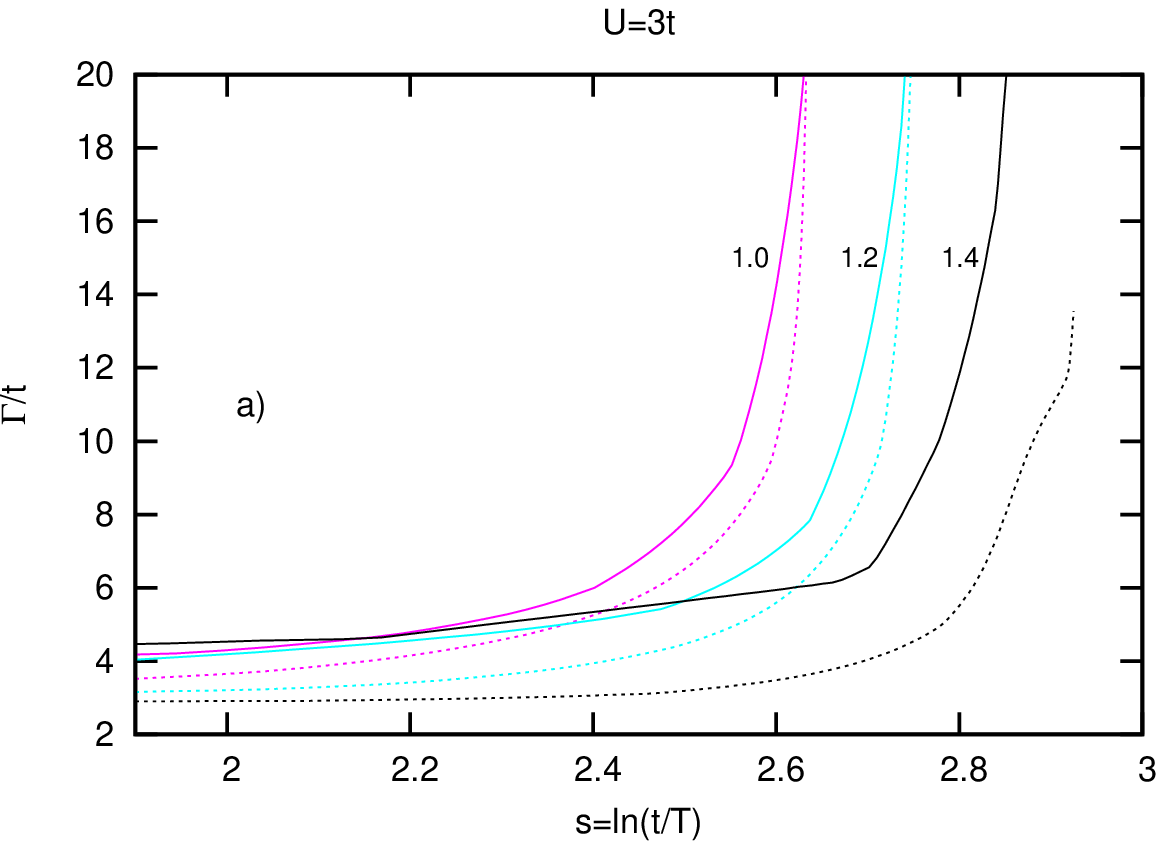}%
\includegraphics[width=0.5\textwidth]{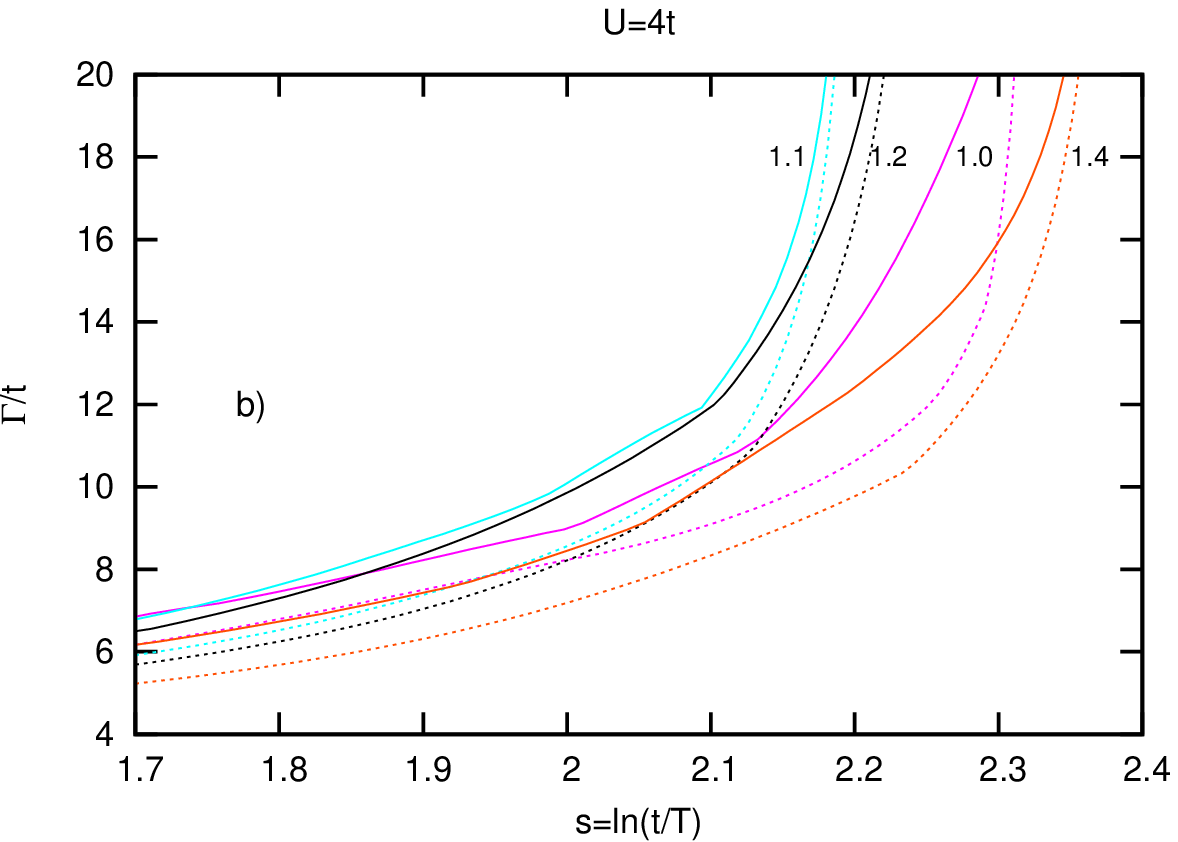}
\caption{The same as in Fig. \protect\ref{vertex_3} for Fermi level being above VHS ($\mu_{\mathrm{eff}}>0$)}
\label{vertex_4}
\end{figure}

The scale dependences of $\Gamma_{\uparrow \downarrow}^{\mathrm{max}}$ and $\Gamma_{\uparrow \downarrow }^{\mathrm{max,E}}$ for different $\bar\mu$ are presented in Fig. \ref{vertex_3} ($\mu_{\rm eff}<0$) and Fig. \ref{vertex_4} ($\mu_{\rm eff}>0$). Let us consider first the case of Fermi level below VHS, $\mu_{\mathrm{eff}}<0$. For both the cases $U=3t$ (Fig. \ref{vertex_3}a) and $U=4t$ (Fig. \ref{vertex_3}b) $\Gamma _{\uparrow \downarrow }^{\mathrm{max}}$ is diverging, and $\Delta\Gamma_{\rm max}/\Gamma _{\uparrow \downarrow }^{\mathrm{max}}$ vanishes or is very small at the end of the flow. Therefore, magnetic fluctuations are predominantly ferromagnetic at the end of the flow. On the other hand, $\Delta\Gamma_{\rm max}$ increases as $\mu_{\rm eff}$ approaches VHS: for both $U=3t$ and $U=4t$ $\Delta\Gamma_{\rm max}$ is the largest in the case $\bar\mu=0.8t$. This means that in the vicinity of VHS the ferromagnetic fluctuations hardly dominate over incommensurate ones. 

For $\mu_{\mathrm{eff}}>0$ (see Fig. \ref{vertex_4}), the difference $\Delta\Gamma_{\rm max}$ is non--zero up to the lowest temperatures where $\Gamma_{\rm max}$ diverges. Moreover, the ratio $\Delta\Gamma_{\rm max}/\Gamma_{\uparrow\downarrow }^{\mathrm{max}}$ increases as $\mu$ increases. We interpret this as that the incommensurate fluctuations are dominating over the ferromagnetic ones. In the case $\bar\mu=1.0t$, where the temperature dependence of $\mu_{\rm eff}(s)$ is strongly non--monotonous (see Fig. \ref{fermi_fig}b), which is possibly related with invalidity of ansatz (\ref{se_ansatz}),(\ref{v_ansatz}) in this regime and causes the shift of the point of vertex diverging to substantially lower temperatures. This case should be considered more correcly in further elaborated studies.

In the next subsection we introduce a small magnetic field to investigate magnetic properties of the system.

%\subsection{Results in finite magnetic field}
\subsection{Magnetization in finite magnetic field}
\label{h<>0}

%\subsubsection{Effective chemical potential}

In this subsection we supplement the picture in zero magnetic field considered above by the results for magnetic response in finite magnetic field. In case $h>0$ we have $\mu_{\mathrm{eff\uparrow}}\neq\mu_{\mathrm{eff\downarrow}}$ due to spin dependence of the spectrum parameters $t_{\mathrm{eff,\sigma}},t_{\mathrm{eff,\sigma}}^{\prime}$ and $\Sigma_{\sigma}$ and, to a small extent, due to the presence of magnetic field. At low scales (high temperatures) the spectrum parameters do not depend on the spin projection substantially and $\mu_{\mathrm{\rm eff\uparrow}}-\mu_{\mathrm{\rm eff\downarrow }}\approx 2h$, but with increasing the scale (lowering temperature) the strong spin dependence can be realized which is a manifestation of exchange enhancement. 

%According to the Mermin--Wagner theorem(??)... However, absolute values of magnetisation are rather small in the end of the flow, which shows that the Mermin--Wagner theorem is fulfilled to a good accuracy.

The strengh of magnetic response is characterized by the magnetization $m$ which can be easily calculated using current parametes of the electronic spectrum $\delta t_{\sigma},\delta t_{\sigma}',\Sigma_{\sigma}$ (we remind that the frequency dependence of the self--energy is neglected),
\begin{equation}
\label{magnetization_def}
m=\frac1{2N}\sum_{\mathbf{k}}\left(f_{\mathbf{k}\uparrow}-f_{\mathbf{k}\downarrow}\right).
\end{equation} 
\begin{figure}
\includegraphics{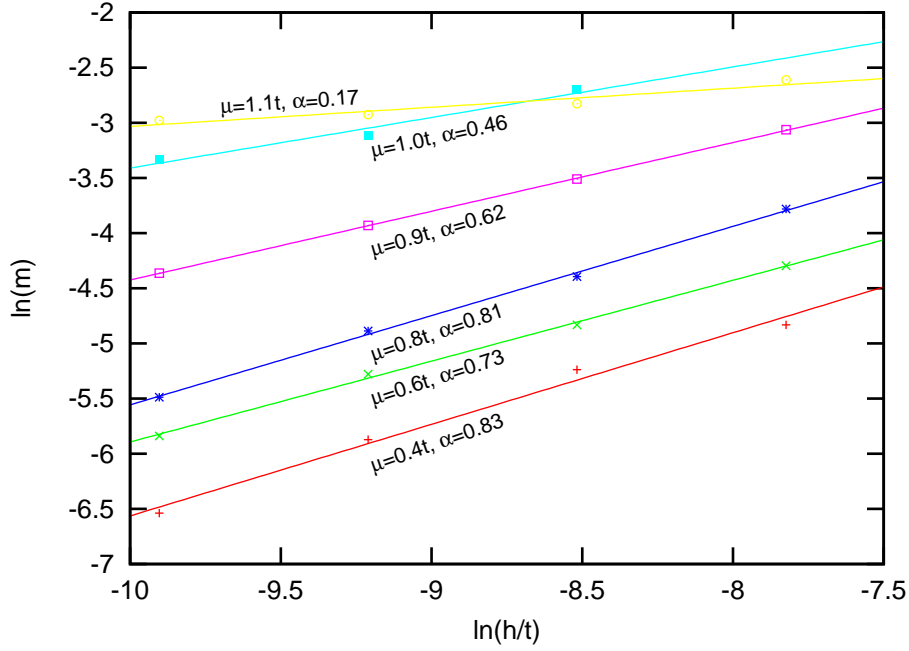} 
\caption{ The logarithmic plots of the dependence $m(h)$ and its fit to $m\propto h^\alpha$. The chemical potentials $\bar\mu$ and fitted $\alpha$ are shown near the plots.}
\label{log}
\end{figure}
\begin{figure}
\includegraphics[width=1.0\textwidth]{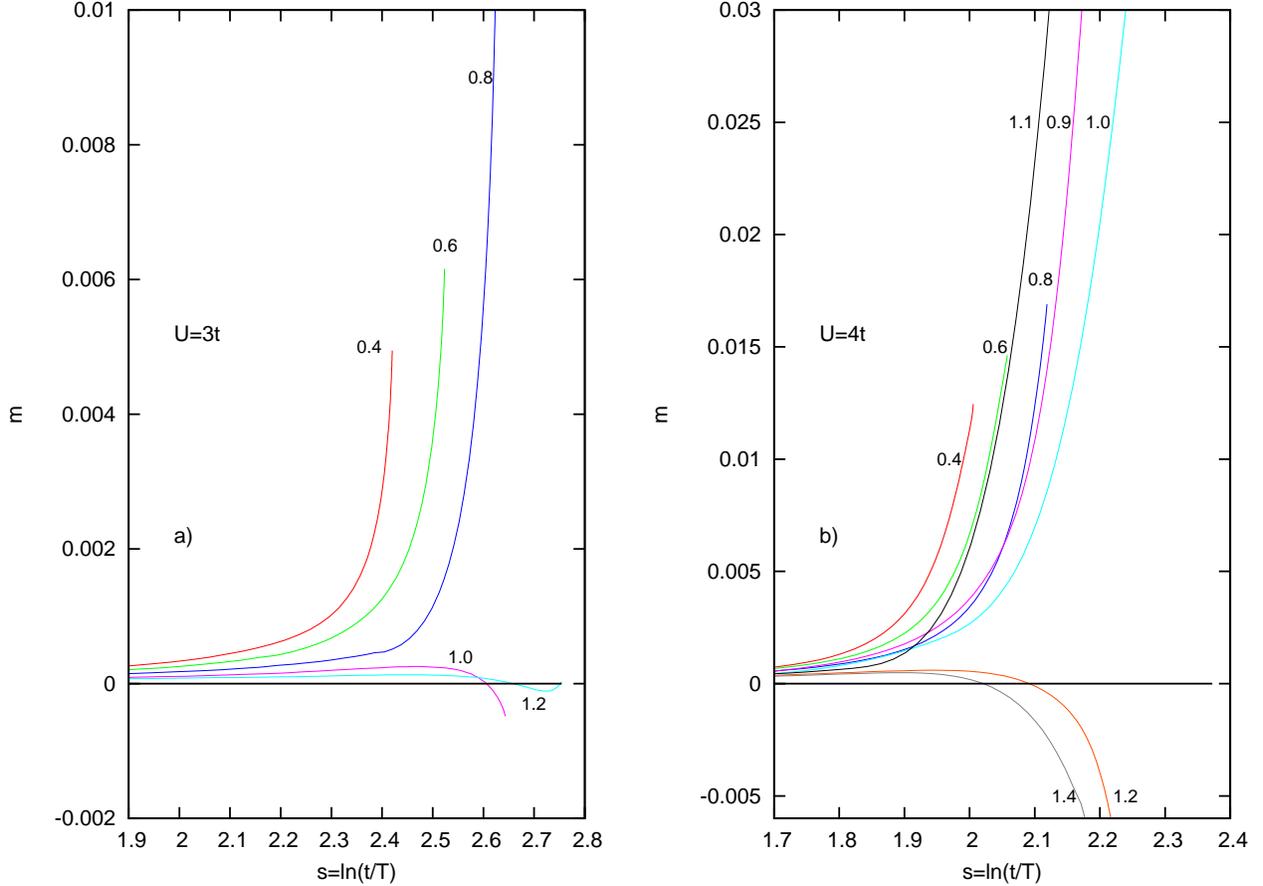}
\caption{The magnetization scale profiles in the field $h=10^{-4}t$ for different chemical potential $\bar\mu$ specified by numbers: a) $U=3t$, b) $U=4t$}
\label{m-s}
\end{figure}
\begin{figure}
\includegraphics[width=1.0\textwidth]{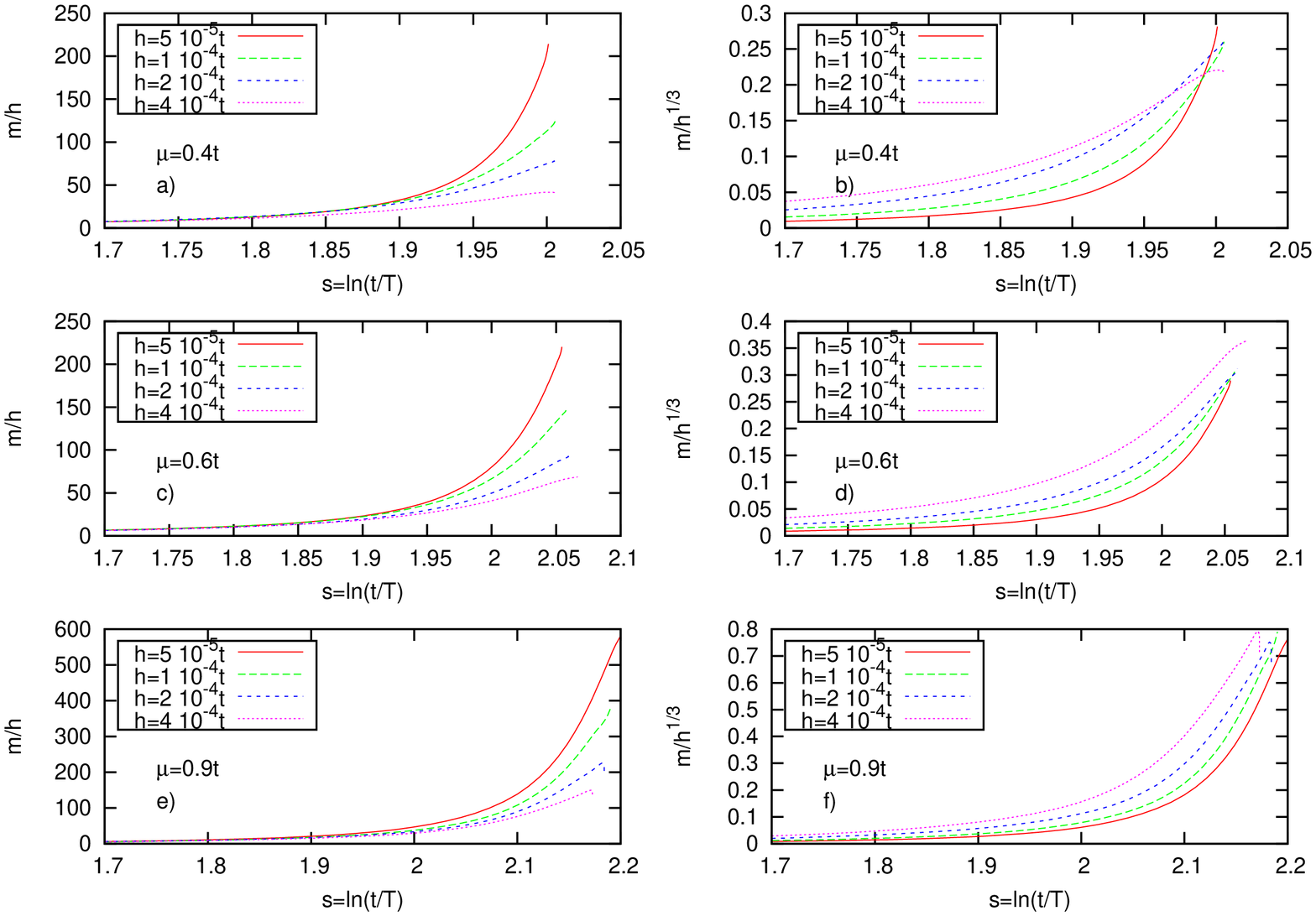}
\caption{The scale profiles of ratios $m/h$ (a,c,e) and $m/h^{1/3}$ (b,d,f) for $U=4t$ at different small magnetic fields: $h/t=(5,10,20,40)\cdot 10^{-5}$ and different bare chemical potentials $\bar\mu$, which values are shown explicitly}
\label{U4_mh_fig}
\end{figure}
Fig. \ref{m-s} shows the comparison of the scale profiles of magnetization $m$ for different bare chemical potentials. For $\mu_{\rm eff}$ above VHS we find slightly negative $m$ at the end of the flow for both the cases $U=3t$ and $U=4t$. This is possibly related to an influence of incommensurate magnetic fluctuations above VHS, as conjectured from the results in zero magnetic field. Below VHS, where dominating ferromagnetic fluctuations were observed in the absence of magnetic field, the magnetic response to magnetic field becomes positive and considerable, especially in the vicinity of VHS: the maximal value of $m$ at the end of the flow increases, but the temperature of sharp increase of the magnetization becomes lower as $\mu$ increases. Note, that the absolute values of magnetization are rather small ($m\ll2n$) at the end of the flow. To verify the fulfillment of Mernin--Wagner theorem, we fit the data for magnetization taken at lowest temperatures of the flow to $m\propto h^\alpha$ dependence (see Fig. \ref{log}), which gives $\alpha\in(0.46,0.83)$ depending on the chemical potential $\mu$. In the vicinity of VHS level ($\bar\mu=1.0t,1.1t$) for $U=4t$ the magnetization is somewhat larger at the end of the flow, in particular $\alpha=0.17$ at $\bar\mu=1.1t$. Additionally, in this case magnetization tends to saturate at lowest temperatures with increasing the magnetic field (not shown). For these chemical potentials the renormalized Fermi level lies very near VHS and the Mermin--Wagner theorem may not be fulfilled to a good accuracy. On the other hand, these cases are on the borderline between the regions of strong ferromagnetic and incommensurate fluctuations. %Therefore, corresponding results should be possibly supplemented by those with a better parametrization of the vertex momentum dependence. 

The temperature dependence of the exponent $\alpha$ is of particular interest, since it allows to investigate the magnetic properties of the system while entering the region of strong magnetic fluctuations. It is obvious that at high temperatures $m\propto h$ ($\alpha=1$), but at low temperatures this relation can be violated due to fluctuations. To determine the temperature of the crossover into the regime with strong ferromagnetic fluctuations, we adopt the criterion $\alpha=1/3$, dictated by the mean--field value of the critical exponent at the conventional magnetic phase transition. We denote the crossover scale as $s^*$ (the corresponding temperature is $T^*=t\exp(-s^*)$). 

Fig. \ref{U4_mh_fig} shows the comparative (with respect to the magnetic field) representation of $m/h$ and $m/h^{1/3}$ scale profiles for different positions of Fermi level below VHS in the case $U=4t$. In cases $\bar\mu=0.4t,0.6t$ the crossover scale is reached within the flow and lies near its end, while in the case $\bar\mu=0.9t$ it is beyond the flow and is obtained by an extrapolation. Therefore, increasing $\bar\mu$ tends to decrease $T^*$. We do not find a well--resolved crossover scale at $U=3t$, but we note that the system is close to it at the end of the flow in the case $\bar\mu=0.4t$, where the magnetic fluctuations tend to be commensurate. This means that at $U=3t$ crossover to ferromagnetic ordering can be realized only at extremely low temperatures which are not available within the present approach.  For the case $\mu_{\mathrm{eff}}>0$ and both $U=3t$ and $U=4t$ we find a relatively weak response which becomes negative close to the end of the flow, as discussed above.

The distribution of temperatures $T_{\rm min}$ and $T^*$ for different chemical potentials yields a natural connection between zero--field ($T_{\rm min}$) and finite--field ($T^*$) results. We find that both $T_{\rm min}$ and $T^*$ decrease as the system approaches VHS, which is possibly an effect of incommensurate magnetic fluctuations and temperature smearing of VHS (above van Hove filling DOS falls down faster than below it); below VHS we obtain $T^*<T_{\rm min}$. However, in the case $\bar\mu=1.1t$ we find $T^*>T_{\rm min}$, which is possibly connected to the abovediscussed violation of the Mermin--Wagner theorem in a close vicinity above VHS.

\subsection{Phase diagram}
\label{Phase_diagram}

\begin{figure}
\includegraphics{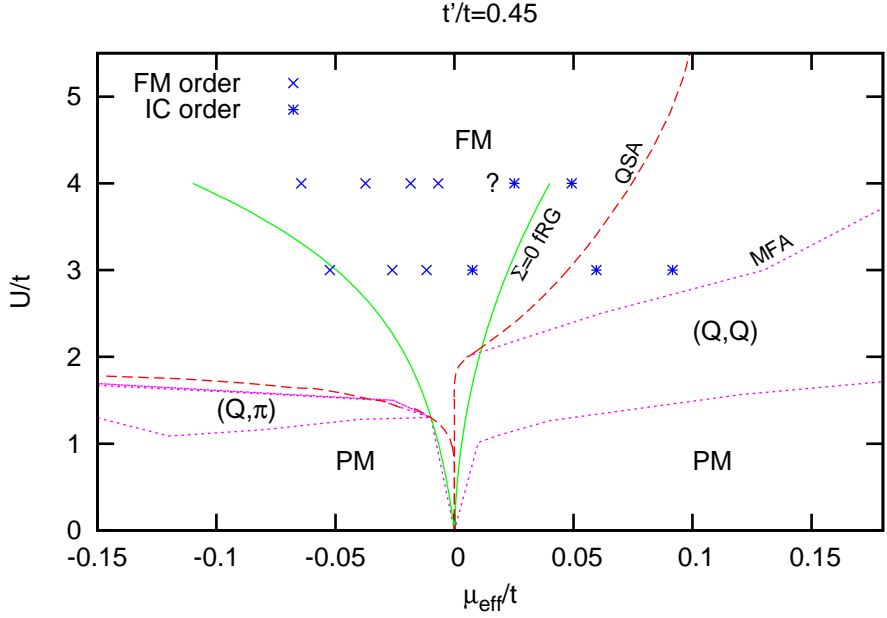} 
\caption{Summary magnetic phase diagram in the vicinity of VH filling in $\protect\mu_{\mathrm{eff}}-U$ plane, $t^{\prime}=0.45t$, obtained within the mean--field \cite{OurIC}(short--dashed lines, $\mu_{\mathrm{eff}}^{\rm MF}=\mu-Un/2$, where $n$ is the electronic density), $\Sigma=0$--fRG \cite{Kampf} (solid lines), QSA \cite{QS} (long--dashed lines) and present SE fRG (symbols) approaches, see text. The boundary lines corresponding to $\Sigma=0$--fRG and QSA approaches separate ferromagnetic and paramagnetic regions below van Hove filling and ferromagnetic and incommensurate magnetic phases above van Hove filling.
}
\label{Summary}
\end{figure}

In this Section we summarize the results obtained in zero (Sect. \ref{h=0}) and finite (\ref{h<>0}) magnetic field and compare our results with the results of previous approaches. The phase diagram constructed in terms of renormalized Fermi level $\mu_{\rm eff}$ (see discussion in Sec. \ref{h=0}) and Coulomb interaction $U$ is shown in Fig. \ref{Summary}. 

In the case $\mu_{\rm eff}<0$ the large region on the phase diagram is found where the zero--field magnetic fluctuations are predominantly commensurate. In finite magnetic field, the system exhibits some indications of ferromagnetic order in the ground state: at $U=3t$ a crossover from paramagnetic to ferromagnetic order is observed well beyond the flow, while at $U=4t$ the crossover to ferromagnetic state is well--resolved in the close vicinity of the end of the flow. In the case $\mu_{\rm eff}>0$ (excluding the region of near vicinity of Fermi level to VHS) we do not find commensurate magnetic fluctuations to be dominating in zero magnetic field: they are competely replaced by incommensurate flucuations with corresponding maximal vertex $\Gamma_{\uparrow \downarrow}^{\mathrm{max}}$ being diverging. This conclusion agrees with the result in finite magnetic field, where the magnetization $m$ becomes slightly negative at low temperatures. Therefore we conclude that ferromagnetic order is suppressed by an incommensurate magnetic fluctuations if the Fermi level is above VHS. The case of $\mu_{\rm eff}$ very near but above VHS is worthy of special attention: in the case $U=3t$ we find incommensurate magnetic fluctuations in zero magnetic field and no indication of ferromagnetic ordering in finite field. However, very near VHS at $U=4t$ we find almost commensurate magnetic fluctuations in zero magnetic field and ferromagnetic--like behavior in finite magnetic field. At the same time, in this case $\mu_{\rm eff}$ depends non--monotonously on $\mu$ and the magnetization on the value of magnetic field. These states are shown by ``?'' symbol in Fig. \ref{Summary}. %Another point is that below VHS the system exhibits strong influence of incommensurate fluctuations as the Fermi level elevates towards VHS. 

The mean--field and quasistatic approximation (MFA and QSA) boundary lines calculated in Refs. \onlinecite{OurIC} and \onlinecite{QS} are shown on the phase diagram in Fig. \ref{Summary} for comparison (the effective chemical potential in MFA $\mu_{\rm eff}^{\rm MF}=\mu-Un/2$ and in QSA it is determined from the condition for electronic density $n(\mu_{\text{eff}})=n_{0}(\mu),$ where $n_{0}(\mu)$ is the number of particles in the non--interacting model). The critical interaction for stability of ferromagnetism within the present approach $U_{\mathrm{c}}>3t$ is somewhat larger then the QSA  and the mean--field resuts ($U_{\mathrm{c}}(\mu_{\mathrm{eff}}=0)\approx2t$),  which naturally suggests that the account for the electronic correlations results in an enhancement of $U_{\mathrm{c}}$. %It is important that QSA yields finite value of $U_{\rm c}$ at VHS. This means that QSA captures an imporatant point that ferromagnetic state can be destroyed by incommensurate magnetic fluctuations. 

The results of the present study partly agree with those calculated within fRG approach without self--energy effects\cite{Kampf} ($\Sigma=0$--fRG approach, see Fig. \ref{Summary}), obtained by studying the temperature dependences of magnetic and superconducting susceptibilities in zero magnetic field. The lower threshold for ferromagnetism $U_{\mathrm{c}}$ obtained in the present study, appears to be finite contrary to $\Sigma=0$--fRG. %At the same time, at VHS this approach yields $U_{\mathrm{c}}=0$. It is interesting that $\Sigma=0$-fRG approach also reveals the presence of narrow region of incommensurate magnetic states above VHS. Its position can be obtained within the approach of Ref. \onlinecite{Katanin} and is shown on the phase diagram (Fig. \ref{Summary}) for $U=4t$. 
At the same time, away from VHS these two approaches qualitatively agree.

The comparison of different approaches MFA, QSA, $\Sigma=0$--fRG, SE fRG shows a step--by--step restriction of the size of ferromagnetic region. Ferromagnetism is practically absent for $\mu_{\rm eff}>0$ within SE fRG, but not restricted with respect to previous approaches in the case $\mu_{\rm eff}<0$. Above VHS ferromagnetism is destroyed by well--resolved incommensurate magnetic fluctuations, while below VHS quantum commensurate fluctuations dominate.

\section{Conclusions}

\label{Conclusions} In this study we present a fRG treatment of magnetic order in the Hubbard model, controlled by the Fermi level being in the vicinity of van Hove singularity (VHS) and on--site Coulomb interaction. The introduced version of fRG accounts for self--energy corrections, which implies a proper account of the Fermi surface moving as the temperature decreases. The following aspects of this problem are investigated: the renormalization of vertices in zero magnetic field, spin splitting in finite field, finite--temperature behavior and the electronic spectrum renormalization.

We find that magnetic properties of the system are substantially asymmetric with respect to the Fermi level position relatively to VHS. In particular, ferromagnetic ordering is strongly suppressed above VHS due to competition with incommesurate magnetic fluctuations. Below VHS we find precurors of ferromagnetic ordering at low temperatures and not too small interaction (of order of half bandwidth). The temperature of the crossover to strong ferromagnetic fluctuations decreases as the Fermi level elevates towards VHS. 

Correlation effects do not change substantially the form of electronic spectrum which is characterized by $t'/t$. However, the renormalized ratio $t'/t$ monotonously increases as the Fermi level rises towards VHS. When the Fermi level is at VHS we find that the ansatz used for the self--energy momentum dependence is too crude to catch delicate features of the renormalization. 

The results of the present approach improve the results of the mean--field approximation\cite{OurIC} and quasistatic approach\cite{QS} and agree with previous fRG study \cite{Kampf} for the position of the Fermi level away from VHS. The Mermin--Wagner theorem is also shown to be fulfilled to a good accuracy in the present approach in the most part of the phase diagram. The magnetic field dependence of magnetization demonstrates power--law behavior with the exponents $\alpha\in(0.62,0.83)$.%Being restricted to quasiparticle picture present approach may not resolve the non--Fermi--liquid contributions, which might be important in some narrow upper vicinity of VHS level.

The observed region of ferromagnetism for the Fermi level above VHS is rather narrow. This explains qualitatively the magnetic behavior of one-- and two--layer ruthenates (which have, after corresponding particle--hole transformation, $\mu_{\rm eff}>0$). The non--Fermi--liquid behavior of doped Sr$_2$RuO$_4$\cite{Kikugawa} can be possibly related to the non--monotonous temperature dependence of the Fermi level near VHS in the present approach, which neglects non--quasiparticle contributions. These contributions may therefore be important for the Fermi level near VHS (see, e.~g. Ref. \onlinecite{Katsnelson}), which  is the subject of future study. The results obtained demonstrate an important role of many--electron renormalizations of electron spectrum, in particular of chemical potential, in the presence of VHS. This fact can be crucial for the criterium and properties of weak itinerant ferromagnetism. Another important point is that peculiarities of ferromagnetic ordering  related to the presence of VHS cannot be even qualitatively captured by any elaborate method based on DOS consideration only. 

The obtained finite--temperature picture, in particular, the relation of the character of magnetic fluctuations to the position of the Fermi level and absence of the long--range order (which is stated by the Mermin--Warger theorem) should be possibly supplemented by a zero-temperature study of this problem. Investigation of the effect of finite (but not too small) magnetic field and possibility of the metamagnetic transitions is also of interest. Another unsolved problem is a unified description of weak and strong ferromagnets, which can receive new insights from weak--coupling investigations of the Hubbard model.

We are grateful to W. Metzner and H. Yamase for stimulating discussions. This work was supported in part by Partnership Program of the Max--Planck Society, the project 'Quantum Physics of Condensed Matter' of Presidium of Russian Academy of Science and President Program of Scientific Schools Support 4711.2010.2.
\appendix

%\section*{Appendix. Numerical treatment of fRG equations}
%In this appendix we discuss the technical details of numerical treatment of fRG equations (\ref{DGamma}),(\ref{DSigma}).

\section{Projecting points}

\label{PP_ansatz} To parametrize the momentum dependence of the self--energy and vertex, we introduce a set of points in the Brillouin zone (below we refer to them as projecting points, PPs), so that the values of functions at these points represent the function. In prevous studies \cite{Salmhofer} this set of the points was chosen as an intersection of lines of constant angles (located at the center of patches) with FS which was assumed to be fixed. Since we account for momentum dependence of the self--energy and vertex more accurately, we supplement this set of PPs at FS by a corresponding set in the vicinity of FS by choosing additional points belonging to ''shifted`` FS and call it auxiliary set of PPs. However, since moving of FS is continuous process, we apply discrete Runge--Kutt procedure (RKP) to solve the system of differential equations numerically, which causes some complications in definition of PP.

At any discrete step of RKP we introduce the PPs of the following types: (i) main PPs, which belong to FS, determined by the chemical potential $\mu$, and the self--energy $\Sigma_{\mathbf{k}\sigma}$ at the beginning of the step. (ii) auxiliary PPs which are detemined analogously to main PPs, with $\mu$ being shifted by $\delta \mu_{\sigma}$ for different spin projections, and $\delta\mu_{\sigma}$ is determined by typical shift of momentum--independent part of self--energy $\Sigma _{\sigma }$ at previous step. (iii) current PPs which are determined by current $\Sigma _{\mathbf{k}\sigma }$ within the step. Note, that at the beginning of the step main PPs coincide with current ones. Introduction of the auxiliary PPs set is needed since, due to definition of main PPs set, the functions chosen to represent momentum dependence of $\Sigma $ ($\cos k_{x}+\cos k_{y},\cos k_{x}\cos k_{y},1$) are linearly dependent at the constant energy set (see Section \ref{self-energy_ansatz}), which does not allow to use them for linear regression of $\Sigma$--derivatives. Therefore, an additional set of PPs is needed; the calculation of the vertex near current FS allows to expand the vertex linearly beyond simple projecting ansatz (see detail explaination in Section \ref{vertex_ansatz}), which was used in earlier studies.

RKP makes its step using the vertex and self--energy derivatives calculated at the intermediate (current) value of argument $s$ at fixed (during the step) sets of main and auxiliary PPs. From this we extract derivatives $\dot{\delta}t_{\sigma}$, $\dot{\delta}t_{\sigma }^{\prime }$ and $\dot{\Sigma}_{\sigma }$, see Section \ref{self-energy_ansatz}. However, the right--hand side of fRG equations contains \textit{current} self--energy function and vertex function, projected on the current FS. RKP allows to calculate this current self--energy function and use it to determine current FS and corresponding current PPs set. Then we apply the procedure of Sect. \ref{vertex_ansatz} accounting for the influence of the FS moving on the current value of projected vertex.


\begin{thebibliography}{99}
\bibitem{Comment} Without introducing approximations, the final rerults should not depend on the choice of $\mu(T)$ dependence provided that the final renormalized chemical potential or particle number is fixed.
\end{thebibliography}
\end{document}